\documentclass[preprint,a4]{elsarticle}
\usepackage{graphicx}
\usepackage[T1]{fontenc}
\usepackage{epsfig}
\usepackage[utf8]{inputenc}
\usepackage[english]{babel}

\journal{Acta Astronautica}

\begin{document}

\begin{frontmatter}

  \title{Objects orbiting the Earth in deep resonance}

\author[rvt1]{J. C. Sampaio\corref{cor1}}
\ead{jarbascordeiro@gmail.com}
\author[rvt2]{E. Wnuk}
\ead{wnuk@amu.edu.pl}
\author[rvt3]{R. Vilhena de Moraes}
\ead{rodolpho.vilhena@gmail.com}
\author[rvt4]{S. S. Fernandes}
\ead{sandro@ita.br}

\cortext[cor1]{Corresponding author. Tel.: +55 1231232800}

\address[rvt1]{UNESP- Univ Estadual Paulista, CEP. 12516-410 Guaratinguetá-SP, Brazil}
\address[rvt2]{AMU, Astronomical Observatory, PL 60-286, Poznan, Poland}
\address[rvt3]{UNIFESP- Univ Federal de São Paulo, 12231-280 São José dos Campos, SP, Brazil}
\address[rvt4]{ITA- Inst Tecnológico de Aeronáutica, 12228-900 São José dos Campos, SP, Brazil}

\renewcommand\refname{\normalsize{References}}

\begin{abstract}
The increasing number of objects orbiting the Earth justifies the great attention and interest in the observation, spacecraft protection and collision avoidance. These studies involve different disturbances and resonances in the orbital motions of these objects distributed by the distinct altitudes. In this work, the TLE (Two-Line Elements) of the NORAD are studied observing the resonant period of the objects orbiting the Earth and the main resonance in the LEO region. The time behavior of the semi-major axis, eccentricity and inclination of some space debris are studied. Possible irregular motions are observed by the frequency analysis and by the presence of different resonant angles describing the orbital dynamics of these objects.
\end{abstract}


\end{frontmatter}

\renewcommand\refname{\normalsize{References}}

\section*{\normalsize 1. Introduction}

The objects orbiting the Earth are classified, basically, in Low Earth Orbit (LEO), Medium Earth Orbit (MEO) and Geostationary Orbit (GEO). Most of the objects are found in the LEO region because this region has a big quantity of space debris. Considering approximately 10000 cataloged objects around the Earth, one can verify the distribution of the objects as: 27$\%$ of operational spacecraft, 22$\%$ of old spacecraft, 41$\%$ of miscellaneous fragments, 17$\%$ of rocket bodies and about 13$\%$ of mission-related objects. The uncatalogued objects larger than 1 cm are estimated in some value between 50000 and 600000 \citep{osiander, ikeda}.

Currently, the orbital motions of the cataloged objects can be analyzed using the 2-line element set of the NORAD (North American Defense) \cite{spacetrack}. The TLE are composed by seven parameters and epoch. These data can be compared, for example, with the model of the orbit propagator situated in the artificial satellite. A similar study is done for the Brazilian satellite CBERS-1 in cooperation with China. In this case, orbital perturbations due to geopotential, atmospheric drag, solar radiation pressure, gravitational effects of the Sun and the Moon are considered in the numerical integration of the orbit and the results are compared with the TLE data \citep{kuga, orlando, space}.

In the last years, the LEO region have been studied about the space debris mitigation due to the increasing number of this kind of object through the years. These aspects englobe the observation, spacecraft protection and collision avoidance \cite{changyin,nishida}. The space debris are composed of aluminum from spacecraft structures, alumina
from solid rocket motor exhausts, zinc and titanium oxides from
thermal control coatings and their size ranges from
several meters to a fraction of a micrometer in diameter \cite{mechishnek}.

The space between the Earth and the Moon has several artificial satellites and distinct objects in some resonance. Synchronous satellites in circular or elliptical orbits have been extensively studied in literature, due to the study of resonant orbits characterizing the dynamics of these satellites since the 60's
\citep{morando,blitzer,garfinkel3,gedeon1,lane,jupp,ely,lima,grosso,diogo,ferreira,sampaio,rossi,neto,deleflie,anselmo,chao,sampaiompe,anselmo2,sampaiocobem}.

In this work, objects in resonant orbital motions are investigated. This search involves all cataloged objects in the TLE files of the NORAD. Figures show regions with objects in deep resonance. The time behavior of the semi-major axis, eccentricity, inclination and the frequency analysis corresponding to the orbital motions of some space debris are studied. Possible irregular motions are observed by the presence of different resonant angles describing the orbital dynamics of these objects.

\section*{\normalsize 2. Resonant objects orbiting the Earth}

In this section, the TLE data are used to verify objects in resonant orbital motions and what resonance has the majority of objects \cite{spacetrack}.

\begin{figure}[h!]
\centering
 \includegraphics[scale=1, width=8.3cm, angle=270]{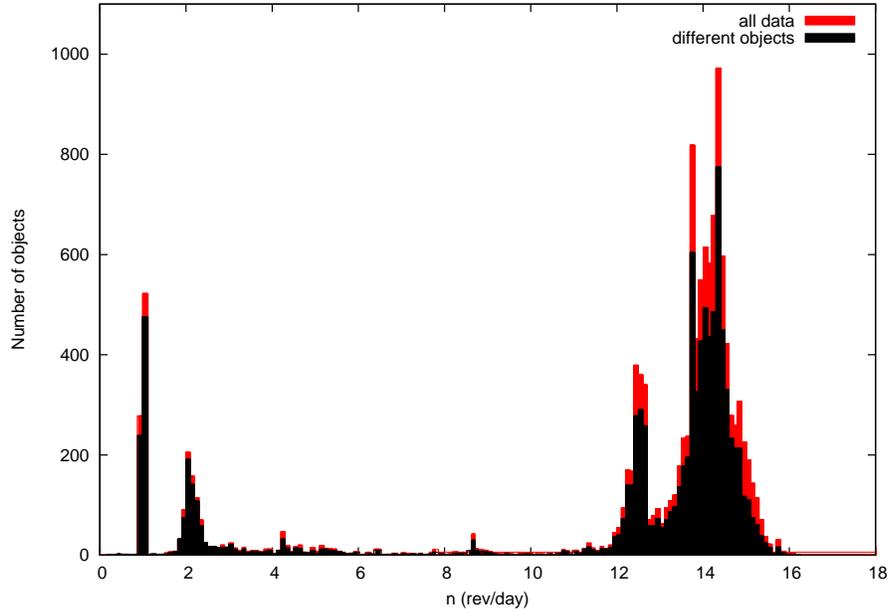}
\caption{Histogram of the mean motion of the cataloged objects.}
\label{histmeanmotion}
\end{figure}

The present distribution of objects by the value of the mean motion $n$ indicates the commensurability between the frequencies of the mean motion of the object and the Earth's rotation motion. See the histogram of the mean motion in Fig. \ref{histmeanmotion}.

 In Fig. \ref{histmeanmotion} it is verified that most of the objects are in the region $13 \leq n (rev/day) \leq 15$.

To study the resonant objects using the TLE data, a criterium is established in the resonant period $Pres$, by the condition $Pres > 300$ days. $Pres$ is obtained by the relation,

\begin{eqnarray}
Pres=\frac{2\pi}{\dot{\phi}_{lmpq}}.
\label{Pres1}
\end{eqnarray}

\noindent{and $\dot{\phi}_{lmpq}$ is calculated from \cite{lane}.

\begin{eqnarray}
 \phi _{lmpq}(M, \omega, \Omega, \Theta)=(l-2p+q)M+(l-2p)\omega +m(\Omega-\Theta-\lambda_{lm})+(l-m)\frac{\pi}{2} \hspace{0.1cm}.
\label{philmpqangle}
\end{eqnarray}

\noindent{with $a$, $e$, $I$, $\Omega$, $\omega$, $M$ are the classical keplerian elements: $a$ is the semi-major axis, $e$ is the eccentricity, $I$ is the inclination of the orbit plane with the equator, $\Omega$ is the longitude of the ascending node, $\omega$ is the argument of pericentre and $M$ is the mean anomaly, respectively; $\Theta$ is the Greenwich sidereal time and $\lambda_{lm}$ is the corresponding reference longitude along the equator. So, $\dot{\phi}_{lmpq}$ is defined as}

\begin{eqnarray}
 \dot{\phi} _{lmpq}=(l-2p+q)\dot{M}+(l-2p)\dot{\omega} +m(\dot{\Omega}-\dot{\Theta}) \hspace{0.1cm}.
\label{phidotres141}
\end{eqnarray}

Substituting $k=l-2p$ in (\ref{phidotres141}),

\begin{eqnarray}
 \dot{\phi} _{kmq}=(k+q)\dot{M}+k\dot{\omega} +m(\dot{\Omega}-\dot{\Theta}) \hspace{0.1cm}.
\label{phidotres1412}
\end{eqnarray}

The terms $\dot{\omega}$, $\dot{\Omega}$ and $\dot{M}$ can be written as \citep{space, justina}.

\begin{eqnarray}
\dot{\omega}=-\frac{3}{4}J_{2}n_{o}\Big{(}\frac{a_{e}}{a_{o}}\Big{)}^{2}\frac{(1-5cos^{2}(I))}{(1-e^{2})^{2}}.
\nonumber
\end{eqnarray}

\begin{eqnarray}
\dot{\Omega}=-\frac{3}{2}J_{2}n_{o}\Big{(}\frac{a_{e}}{a_{o}}\Big{)}^{2}\frac{(cos(I))}{(1-e^{2})^{2}}.
\nonumber
\end{eqnarray}

\begin{eqnarray}
\dot{M}=n_{o}-\frac{3}{4}J_{2}n_{o}\Big{(}\frac{a_{e}}{a_{o}}\Big{)}^{2}\frac{(1-3cos^{2}(I))}{(1-e^{2})^{3/2}}.
\label{omdotOmdotMdot}
\end{eqnarray}

\noindent{$a_{e}$ is the Earth mean equatorial radius, $a_{e}$=6378.140 \emph{km}, $J_{2}$ is the second zonal harmonic, $J_{2}=1,0826\times 10^{-3}$.}

The term $\dot{\Theta}$ in $rad/day$ is

\begin{eqnarray}
\dot{\Theta} \approx 1.00273790926\times2\pi.
\label{omdotOmdotMdotThetadot}
\end{eqnarray}

 In order to use orbital elements compatible with the way in which Two-Line Elements were generated, some corrections are done in the mean motion of the TLE data. Considering as $n_{1}$ the mean motion of the 2-line, the semi-major axis $a_{1}$ is calculated \cite{space}.

\begin{eqnarray}
a_{1}=\left(\frac{\sqrt{\mu}}{n_{1}}\right)^{2/3}
\label{a1fromno}
\end{eqnarray}

\noindent{where $\mu$ is the Earth gravitational parameter, $\mu$=3.986009 x 10$^{14}$ $m$$^{3}$/$s^{2}$.} Using $a_{1}$, the parameter $\delta_{1}$ is calculated by the Eq. (\ref{a1fromnodelta1}) \cite{space}.

\begin{eqnarray}
\delta_{1}=\frac{3}{4}J_{2}\frac{a^{2}_{e}}{a^{2}_{1}}\frac{(3cos^{2}(I)-1)}{(1-e^{2})^{3/2}},
\label{a1fromnodelta1}
\end{eqnarray}

Now, the new semi-major axis $a_{o}$ used in the calculations of the resonant period is defined using $\delta_{1}$ from the Eq. (\ref{a1fromnodelta1}) \cite{space}.

\begin{eqnarray}
a_{o}=a_{1}\left[1-\frac{1}{3}\delta_{1}-\delta^{2}_{1}-\frac{134}{81}\delta^{3}_{1}\right].
\label{a1fromnodelta1a}
\end{eqnarray}

\noindent{and the new mean motion $n_{o}$ used in the calculations is found considering the semi-major axis corrected $a_{o}$}

\begin{eqnarray}
n_{o}=\sqrt{\frac{\mu}{a^{3}_{o}}}.
\label{a1fromnodelta1a}
\end{eqnarray}

\newpage
Now, using the corrections previously shown, a TLE file is studied with the purpose to investigate what resonance (defined by the commensurability between the mean motion of the object and the Earth's rotation angular velocity) has most of the objects orbiting the Earth. After this analysis some space debris in deep resonance are investigated in the third section.

The file analyzed is "$alldata\_2011\_045$" from the website of the Space Track \cite{spacetrack}, and it corresponds to february 2011. In Tab. 1, the number of data and the number of different objects are specified.

\begin{Large}
\begin{table}[h!]
\caption{2-line data of objects orbiting the Earth. }
\centering
\begin{tabular}{ c c c}
\hline File & Number of data & Number of different objects     \\
\hline $alldata\_2011\_045$ & 15360 & 9745 \\ \hline
\end{tabular}
\end{table}
\end{Large}

The simulation identified objects with resonant period greater than 300 days. Several values of the coefficients, $k$, $q$ and $m$ are considered in the Eq. (\ref{philmpqangle}) producing different resonant angles to be analyzed by the Eq. (\ref{Pres1}). See Tab. 2 showing details about the results of the considered simulation.

\begin{Large}
\begin{table}[h!]
\caption{Results of the simulation: objects in resonant orbital motions }
\centering
\begin{tabular}{c c c c c}
\hline Number of data & Number of different objects & coefficient k &  coefficient q & coefficient m    \\
\hline 3180  & 1276 & $-50 \leq k \leq 50$ & $-5 \leq q \leq 5$ & $1 \leq m \leq 50$   \\ \hline
\end{tabular}
\end{table}
\end{Large}

Note that Tab. 2 shows only the data satisfying the established criterium $Pres > 300$ days. The results show the resonant angles and resonant periods which compose the orbital motions of the resonant objects.

Figs. \ref{histmeanmotion2} and \ref{histmeanmotion3} show the semi-major axis versus eccentricity and the semi-major axis versus inclination, respectively, using only the data of the related objects in Tab. 2.

\begin{figure}[h!]
\centering
 \includegraphics[scale=1.6, width=12cm, angle=0]{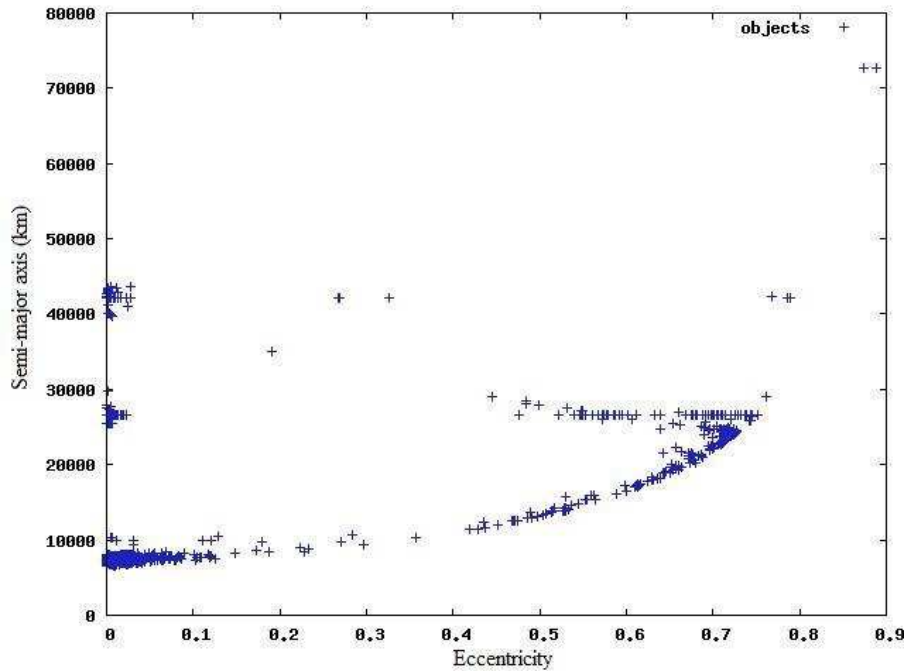}
\caption{Semi-major axis versus eccentricity of resonant objects satisfying the condition of the $Pres > 300$ days.}
\label{histmeanmotion2}
\end{figure}

\vspace{0.3cm}
\begin{figure}[h!]
\centering
 \includegraphics[scale=1.6, width=12cm, angle=0]{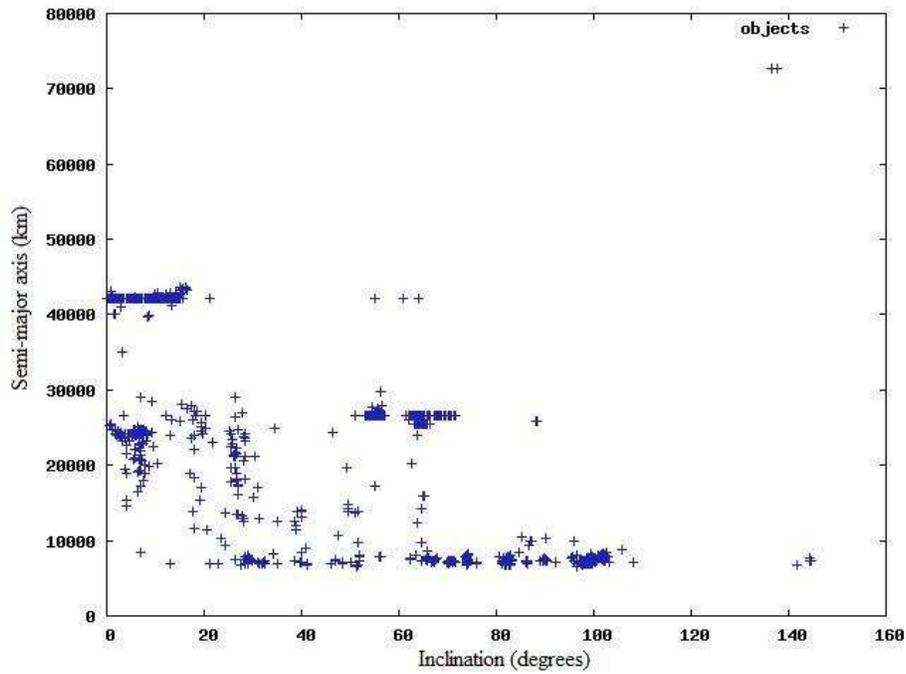}
\caption{Semi-major axis versus inclination of resonant objects satisfying the condition of the $Pres > 300$ days.}
\label{histmeanmotion3}
\end{figure}

\newpage

The objects with Pres > 300 days can be visualized around the Earth in three different regions by the value of the semi-major axis, $a < 15000 km$, $15000 km \leq a < 40000 km$ and $a \geq 40000 km$, see Fig. \ref{fig3dmotionEarth}.

\begin{figure}[h!]
\centering
 \includegraphics[scale=1.6, width=12cm, angle=0]{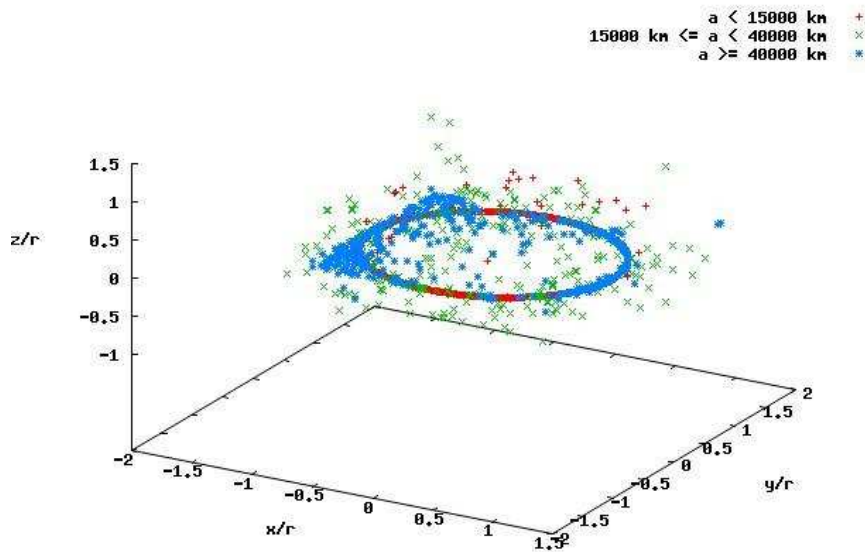}
\caption{Objects around the Earth satisfying the condition of the resonant period greater than 300 days.}
\label{fig3dmotionEarth}
\end{figure}

\noindent{Tab. \ref{tabmres3} shows values of the coefficient $m$ and the number of objects. It is possible to verify that most of the orbital motions of objects are related with the coefficient $m=14$ considering the value of the semi-major axis up to $15000 km$.  }

\begin{normalsize}
\begin{table}[h!]
\caption{Number of objects by the number of the coefficient $m$ considering the value of the semi-major axis up to $15000 km$ }
\centering
\begin{tabular}{ c c c c}
\hline $m$ & \small{Number of objects} &  $m$ & \small{Number of objects} \\
\hline 7 & 2 & 30 & 31   \\ \hline 8 & 1 & 31 & 13  \\ \hline 9 & 3 & 32 & 5  \\ \hline 10 & 1 & 33 & 2  \\ \hline 11 & 3 & 34 & 1  \\ \hline 12 & 20 & 35 & 4 \\ \hline 13 & 31 & 36 & 5  \\ \hline 14 & \textbf{351} & 37 & 42  \\ \hline 15 & 65 & 38 & 14  \\ \hline 16 & 4 & 39 & 14 \\ \hline 18 & 3 & 40 & 24   \\ \hline 20 & 1 & 41 & 186  \\ \hline 21 & 2 & 42 & 116   \\ \hline 22 & 2& 43 & 114   \\ \hline 23 & 2 & 44 & 52  \\ \hline 24 & 10 & 45 & 25  \\ \hline 25 & 93 & 46 & 8  \\ \hline 26 & 19 & 47 & 4   \\ \hline 27 & 66 & 48 & 6  \\ \hline 28 & 179 & 49 & 22   \\ \hline 29 & 97 & 50 & 33 \\ \hline
\end{tabular}
\label{tabmres3}
\end{table}
\end{normalsize}

Comparing the number of objects satisfying the condition of $Pres > 300$ days in the different regions,  $a < 15000 km$ and $a \geq 15000 km$, it is verified that about $62.37 \%$ of the resonant objects has $a < 15000 km$. See this information in Tabs. \ref{tabcomp1} and \ref{tabcomp2}.

\begin{Large}
\begin{table}[h!]
\caption{Number of objects satisfying the condition Pres > 300 days}
\centering
\begin{tabular}{ c c c c}
\hline $a < 15000 km$ & $15000 km \leq a < 40000 km$ &  $a \geq 40000 km$    \\
\hline 1276 &  331 & 439    \\ \hline
\end{tabular}
\label{tabcomp1}
\end{table}
\end{Large}

\begin{Large}
\begin{table}[h!]
\caption{Percentage of objects satisfying the condition Pres > 300 days}
\centering
\begin{tabular}{ c c c c}
\hline $a < 15000 km$ & $15000 km \leq a < 40000 km$ &  $a \geq 40000 km$    \\
\hline 62.37 \% &  16.18 \% & 21.46 \%   \\ \hline
\end{tabular}
\label{tabcomp2}
\end{table}
\end{Large}

These studies allow to investigate the real influence of the resonance effect in the orbital dynamics of the artificial satellites and space debris. The number of resonant objects in comparison with the total number of objects in the TLE data shows the great influence of the commensurability between the mean motion of the object and the Earth's rotation angular velocity in its orbits. In this way, a more detailed study about the resonant period and the resonant angles is necessary.

In the next section, the orbital motions of some space debris in deep resonance are studied.

\newpage
\section*{\normalsize 3. Study of objects in deep resonance}

 Considering the results of the simulations shown in the second section, some cataloged space debris are studied with respect to the time behavior of the semi-major axis, eccentricity, inclination, resonant period, resonant angle and the frequency analysis of the orbital elements. These data are analyzed in this section observing the possible regular or irregular orbital motions.

Fig. \ref{histmeanmotion6} shows the semi-major axis versus resonant period of objects satisfying the condition $Pres > 300$ days. The objects are distributed by the value of semi-major axis in three different regions: 1) $a < 15000 km$, 2) $ 15000 km \leq a < 40000 km$ and 3) $a \geq 40000 km$.

\begin{figure}[h!]
\centering
 \includegraphics[scale=1.2, width=9cm, angle=270]{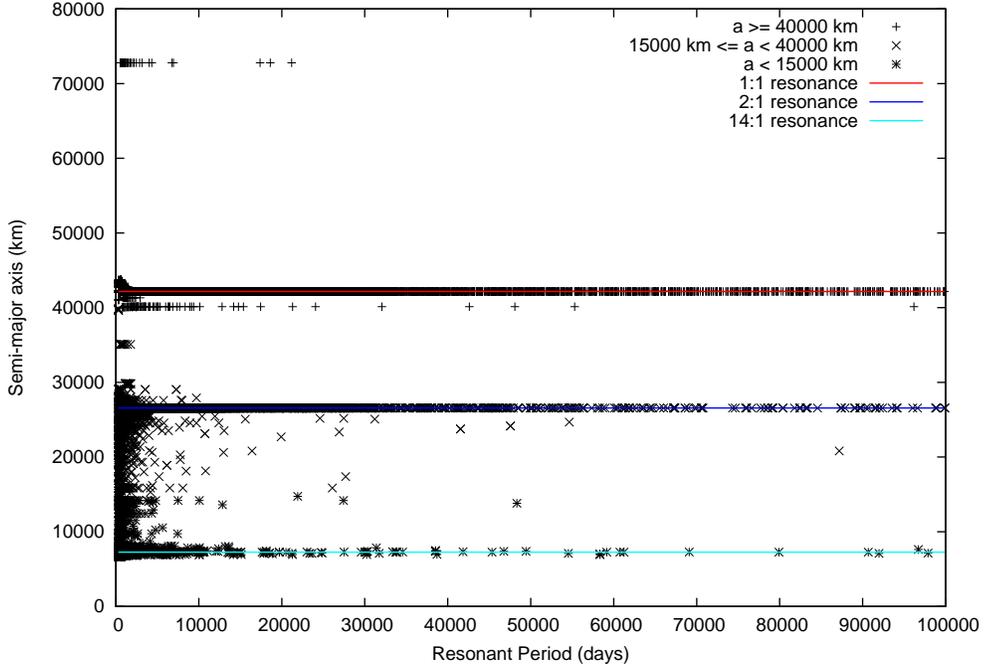}
\caption{Semi-major axis versus resonant period of objects satisfying the condition $Pres > 300$ days.}
\label{histmeanmotion6}
\end{figure}

Analyzing Fig. \ref{histmeanmotion6}, it is possible to verify some objects in deep resonance, in other words, the orbital motions of these objects are influenced by the resonance for a long time. In order to study the orbital motions of these objects, four cataloged space debris are analyzed. From the TLE data, objects are identified by the numbers 325, 546, 2986 and 4855. These objects were chosen because they show resonant period $Pres > 10000$ days.

Figs. \ref{timebehandfreqaviorobj32501} to \ref{timebehandfreqaviorobj485503} show the time behavior of the semi-major axis, eccentricity and inclination of the objects 325, 546, 2986 and 4855, and the frequency analysis of these orbital elements.

The frequency analysis is used as a tool to verify the chaotic orbits. Several authors, \citep{powell,laskar,milos,callegarijr,callegarijr2,ferraz}, also use this method to study the possible regular or irregular orbits in different dynamical systems. A brief description of the frequency analysis involving the FFT (Fast Fourier Transforms), described with more details in \citep{powell, laskar, ferraz}, is presented in what follows.

Generally, the frequency analysis is applied in the output of the numerical integration in the study of dynamical systems. However, in the present work, this analysis is used in the real data of the orbital motions of the objects 325, 546, 2986 and 4855, from the TLE data of the NORAD \cite{spacetrack}.

For regular motions, the orbital elements $oe(t)$ show a dependence on time as follows \citep{powell, laskar, ferraz}:

\begin{eqnarray}
oe(t)=\sum_{h} A_{h} e^{2 \pi i {\textbf{h}} {\textbf{f}} t}
\label{deffreqanalysis}
\end{eqnarray}

\noindent{where $A_{h}$ represent the amplitudes, \textbf{h} $\in$ $\textbf{Z}^{N}$ and \textbf{f} is a frequency vector. The components of \textbf{f} compose the fundamental frequencies of motion, and the spectral decomposition of the orbital motion is obtained from the Fourier transform when the independent frequencies are constant in the course of time \cite{ferraz}. It is possible to observe that \textbf{f} depends on the kind of trajectory \cite{powell}. }

The same numerical procedure is done for irregular and regular motions. The difference in behavior of these orbital motions is used to identify the two kinds of trajectories. The power spectrum, produced by the Fourier transform, of regular motion can be distinguished because it generally has a small number of frequency components. The irregular trajectories are not conditionally periodic and they do not compose an invariant tori. In this way, the Fourier transform of an orbital element, for example, is not a sum over Dirac $\delta$-functions, and consequently the power spectrum is not discrete for irregular motions \citep{powell,ferraz}.

\begin{figure}[h!]
\begin{center}
	\includegraphics[width=6.7cm,height=8cm, angle=270]{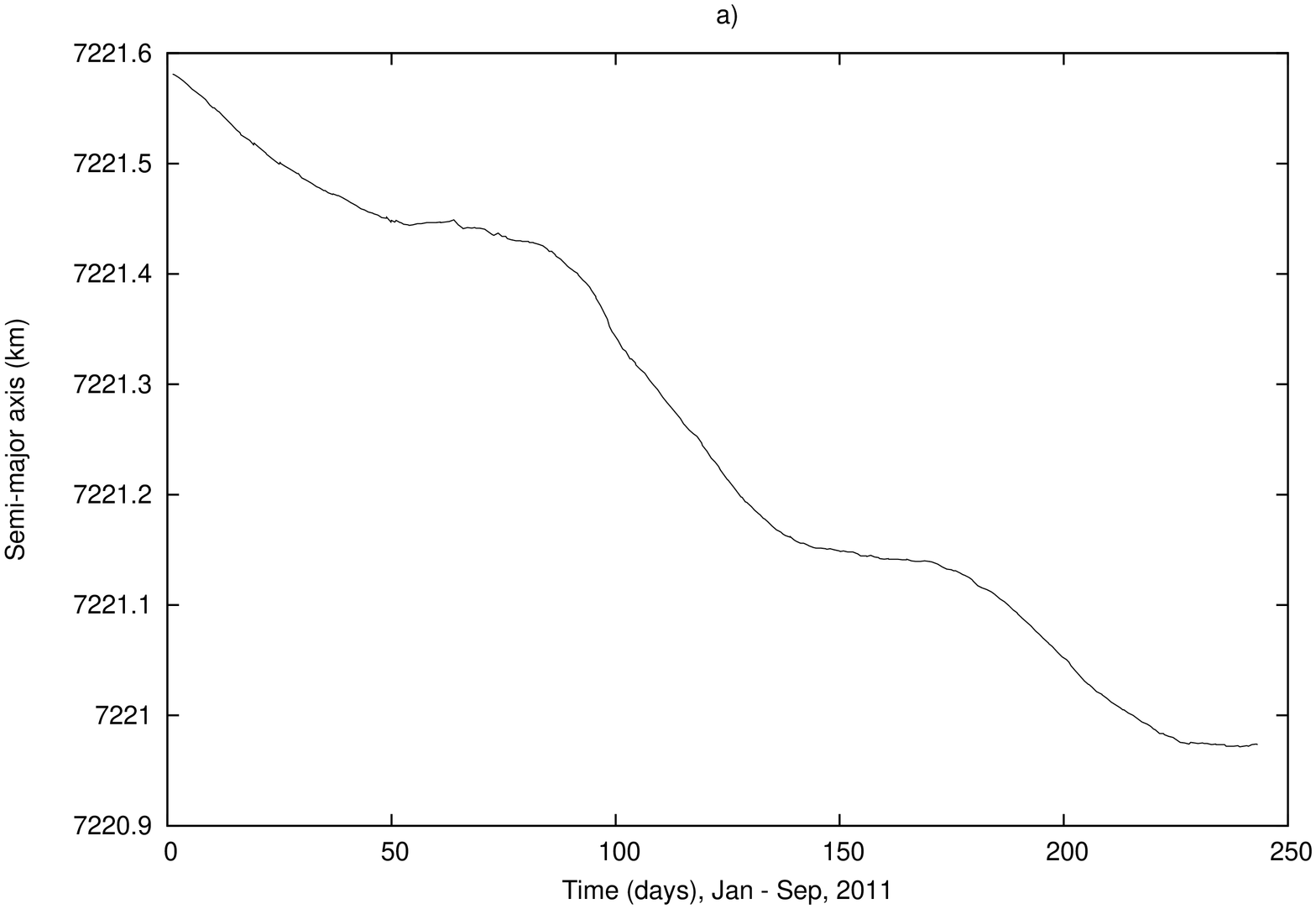} \quad
	\includegraphics[width=6.7cm,height=7.5cm, angle=270]{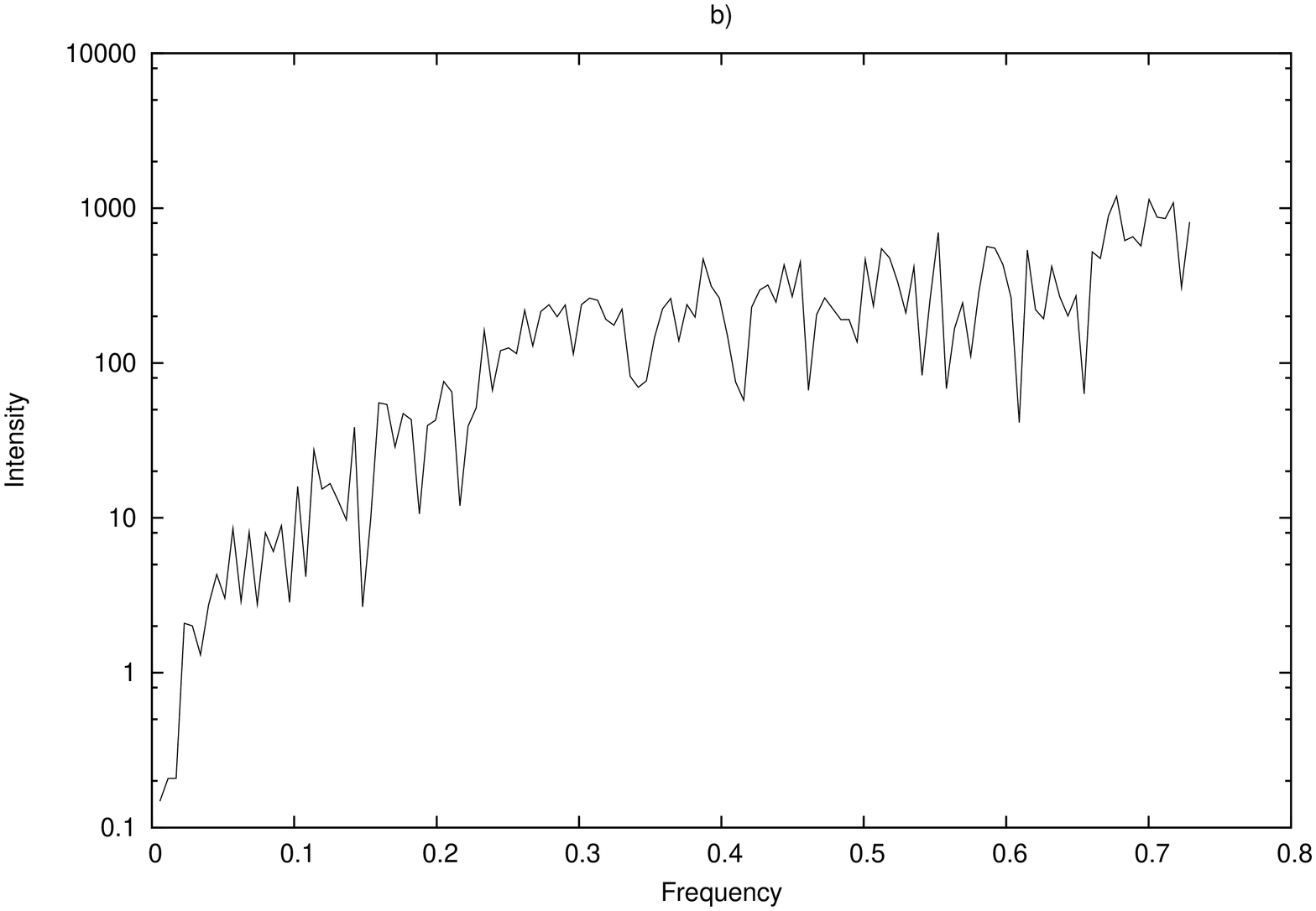}
\caption{Orbital motion of object 325 corresponding to January - September, 2011; a) Time behavior of the semi-major axis and b) Power spectrum of the semi-major axis.} \label{timebehandfreqaviorobj32501}
\end{center}
\end{figure}

\begin{figure}[h!]
\begin{center}
	\includegraphics[width=6.7cm,height=8cm, angle=270]{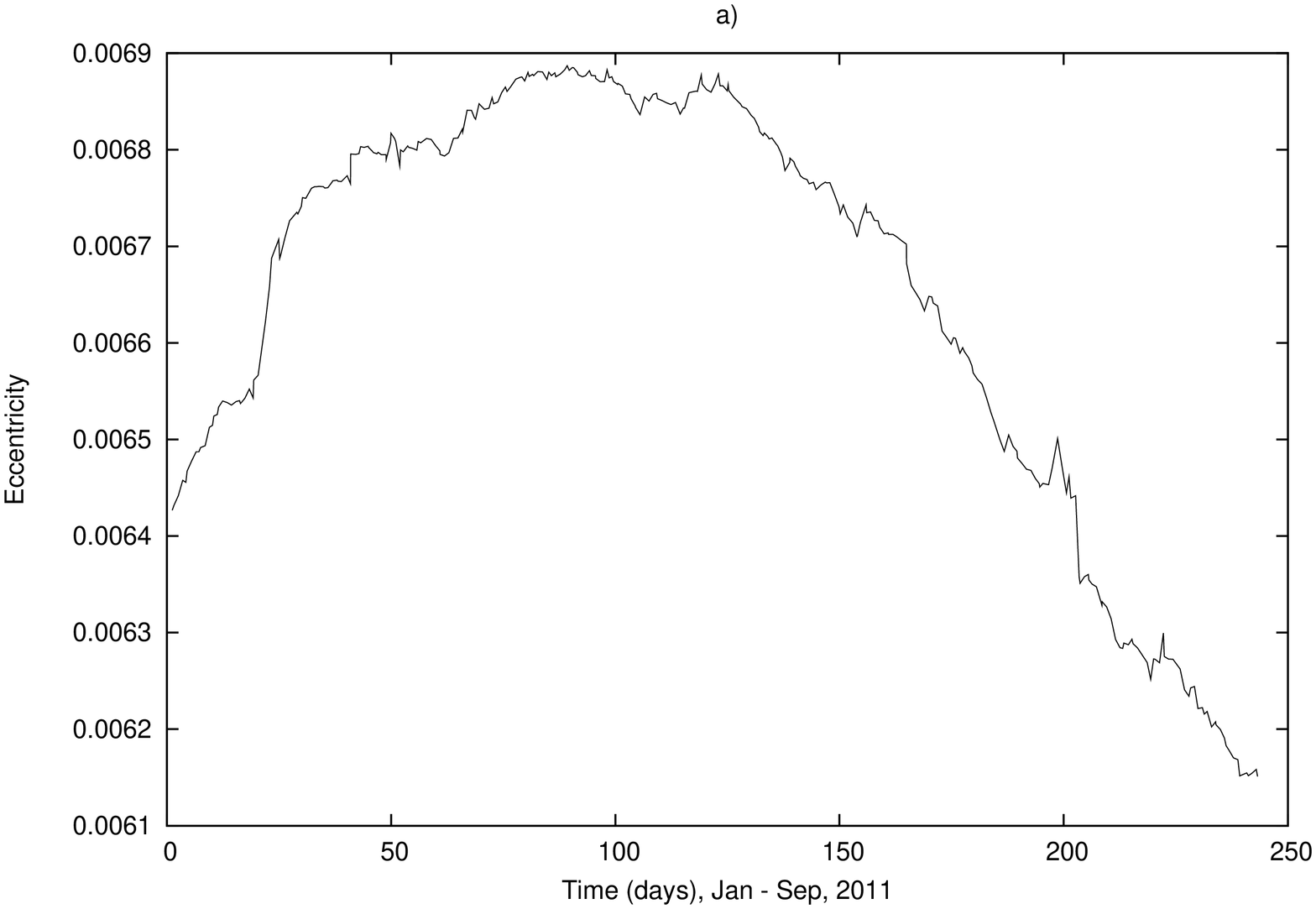} \quad
	\includegraphics[width=6.7cm,height=7.5cm, angle=270]{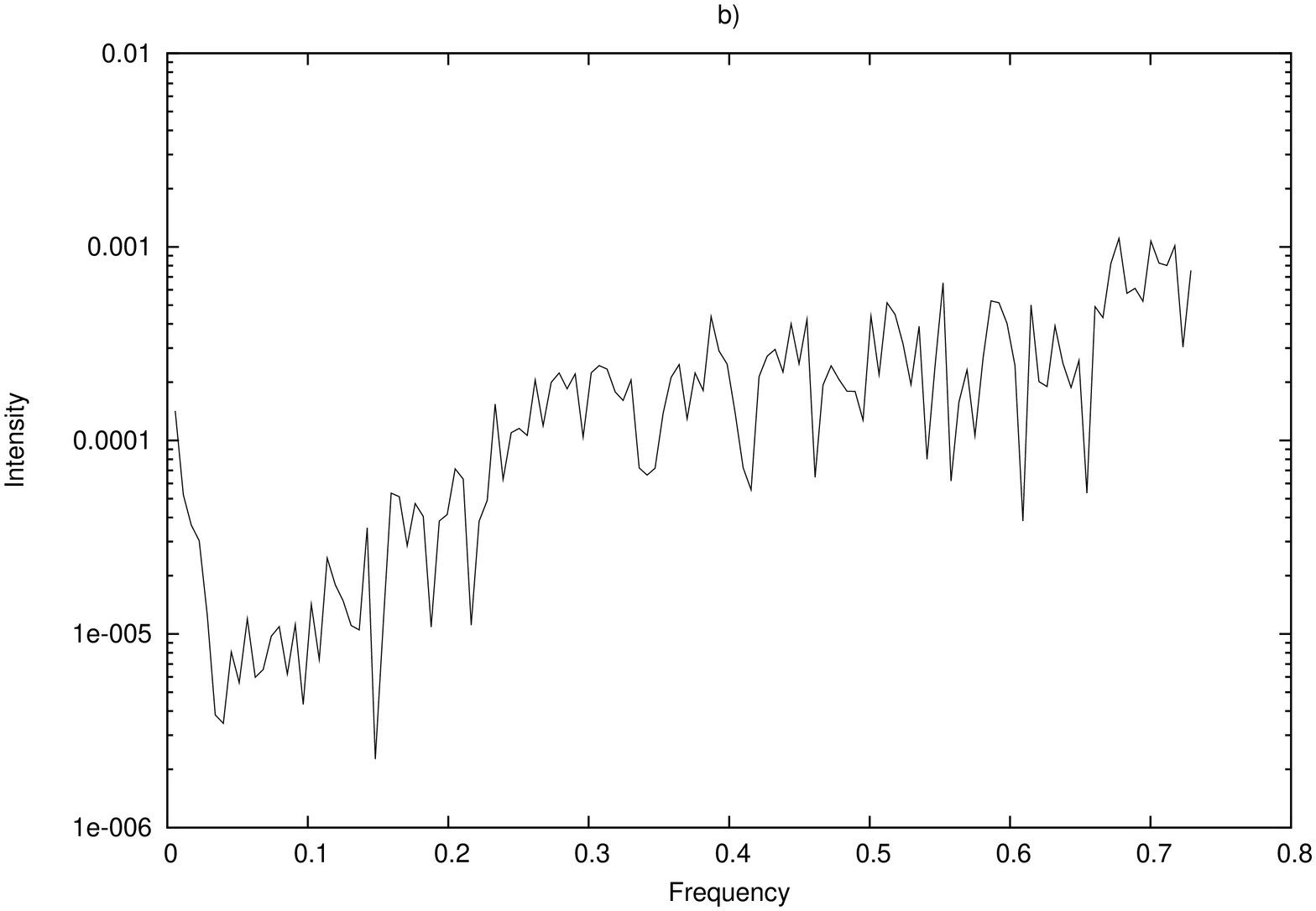}
\caption{Orbital motion of object 325 corresponding to January - September, 2011; a) Time behavior of the eccentricity and b) Power spectrum of the eccentricity.} \label{timebehandfreqaviorobj32502}
\end{center}
\end{figure}

\newpage
\begin{figure}[h!]
\begin{center}
	\includegraphics[width=7cm,height=8cm, angle=270]{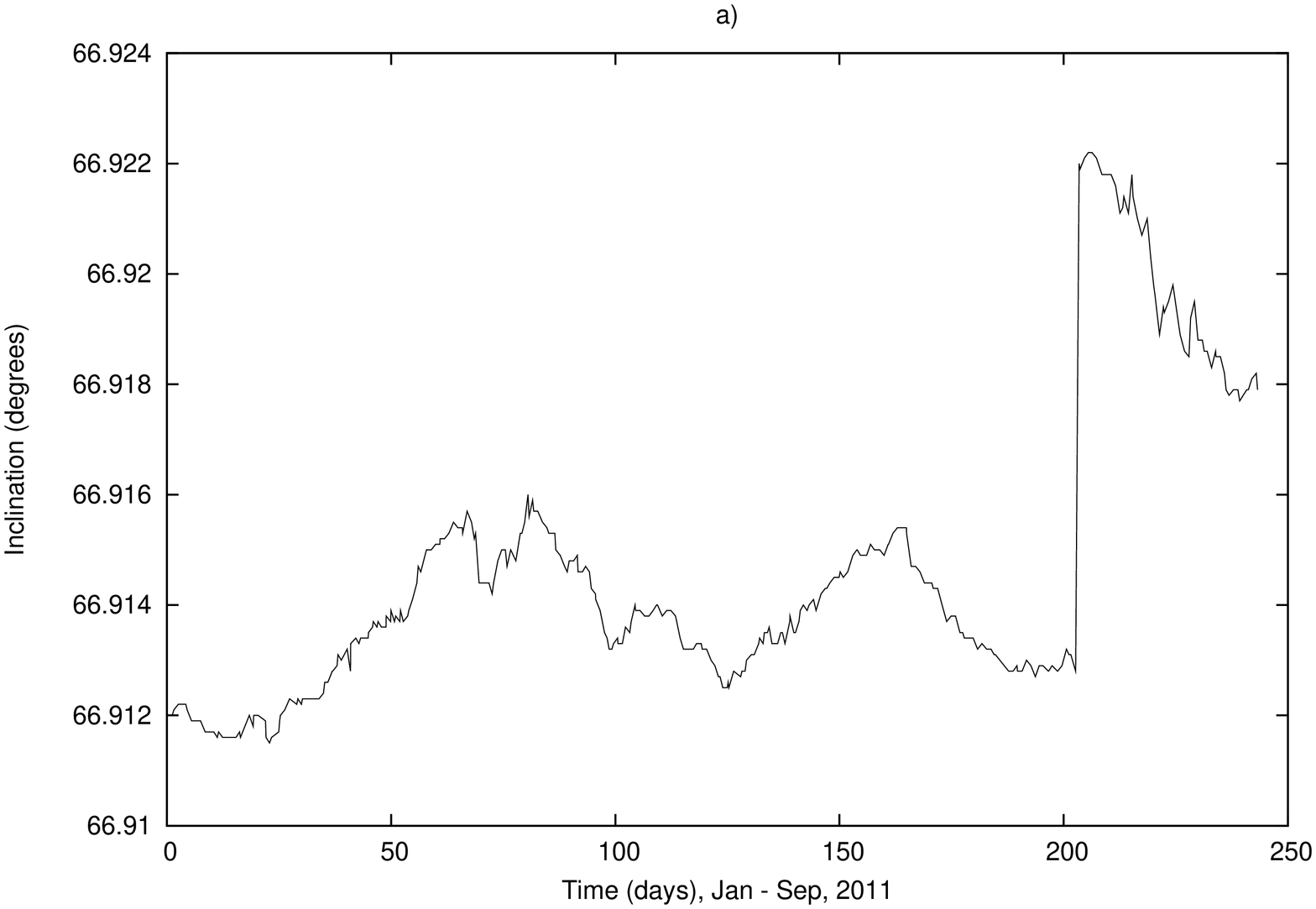} \quad
	\includegraphics[width=7cm,height=7.5cm, angle=270]{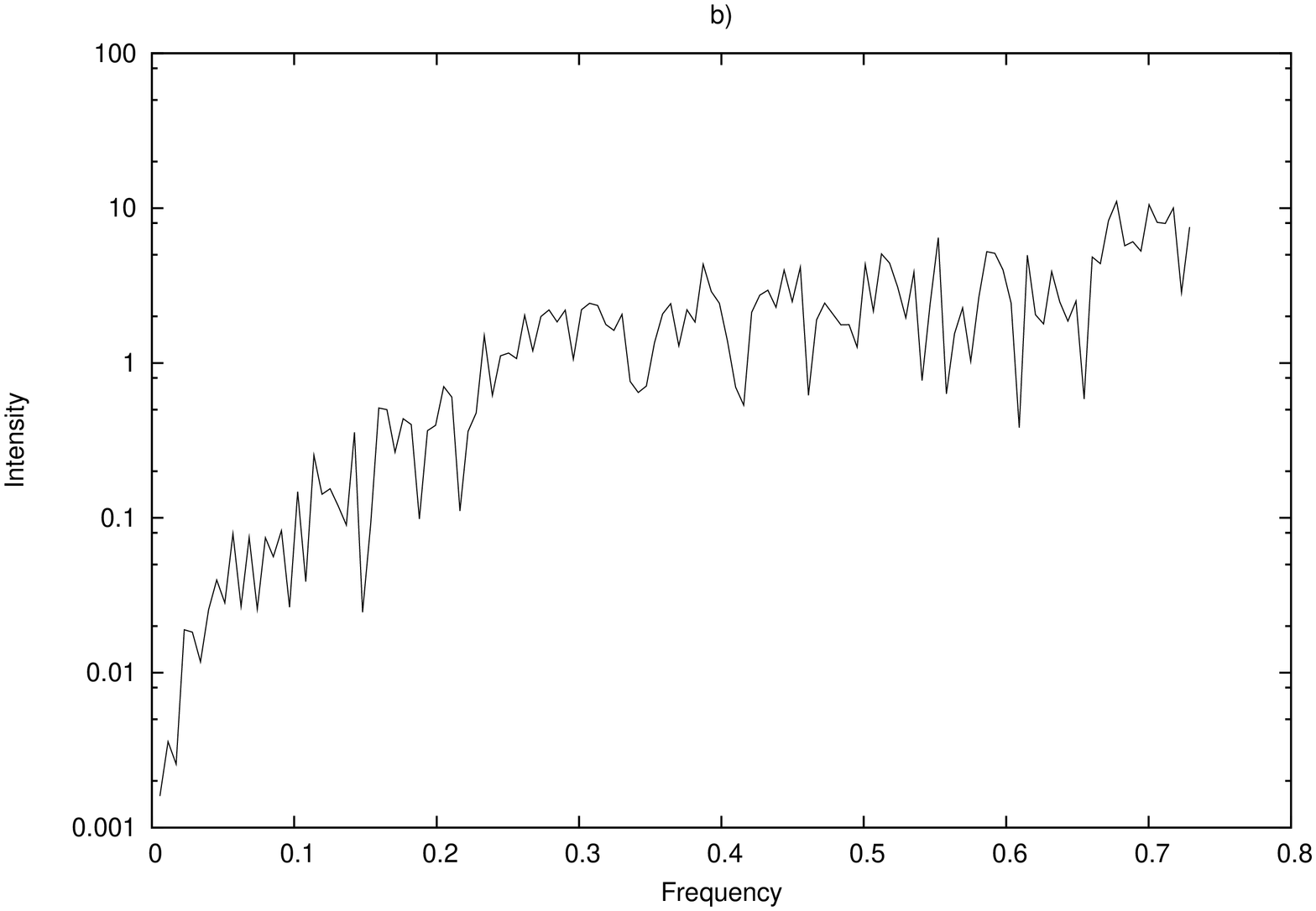}
\caption{Orbital motion of object 325 corresponding to January - September, 2011; a) Time behavior of the inclination and b) Power spectrum of the inclination.} \label{timebehandfreqaviorobj32503}
\end{center}
\end{figure}

\begin{figure}[h!]
\begin{center}
	\includegraphics[width=7cm,height=8cm, angle=270]{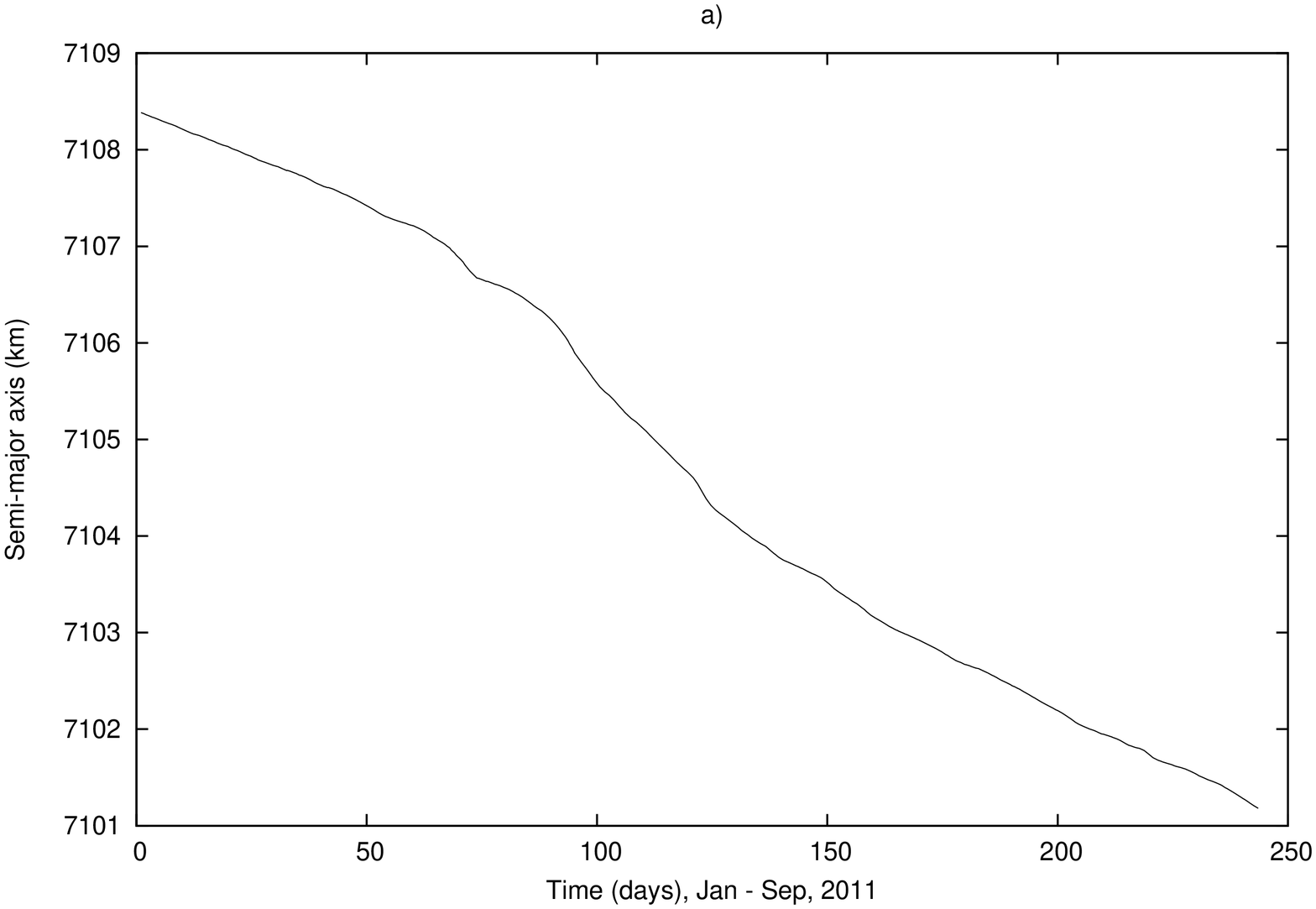} \quad
	\includegraphics[width=7cm,height=7.5cm, angle=270]{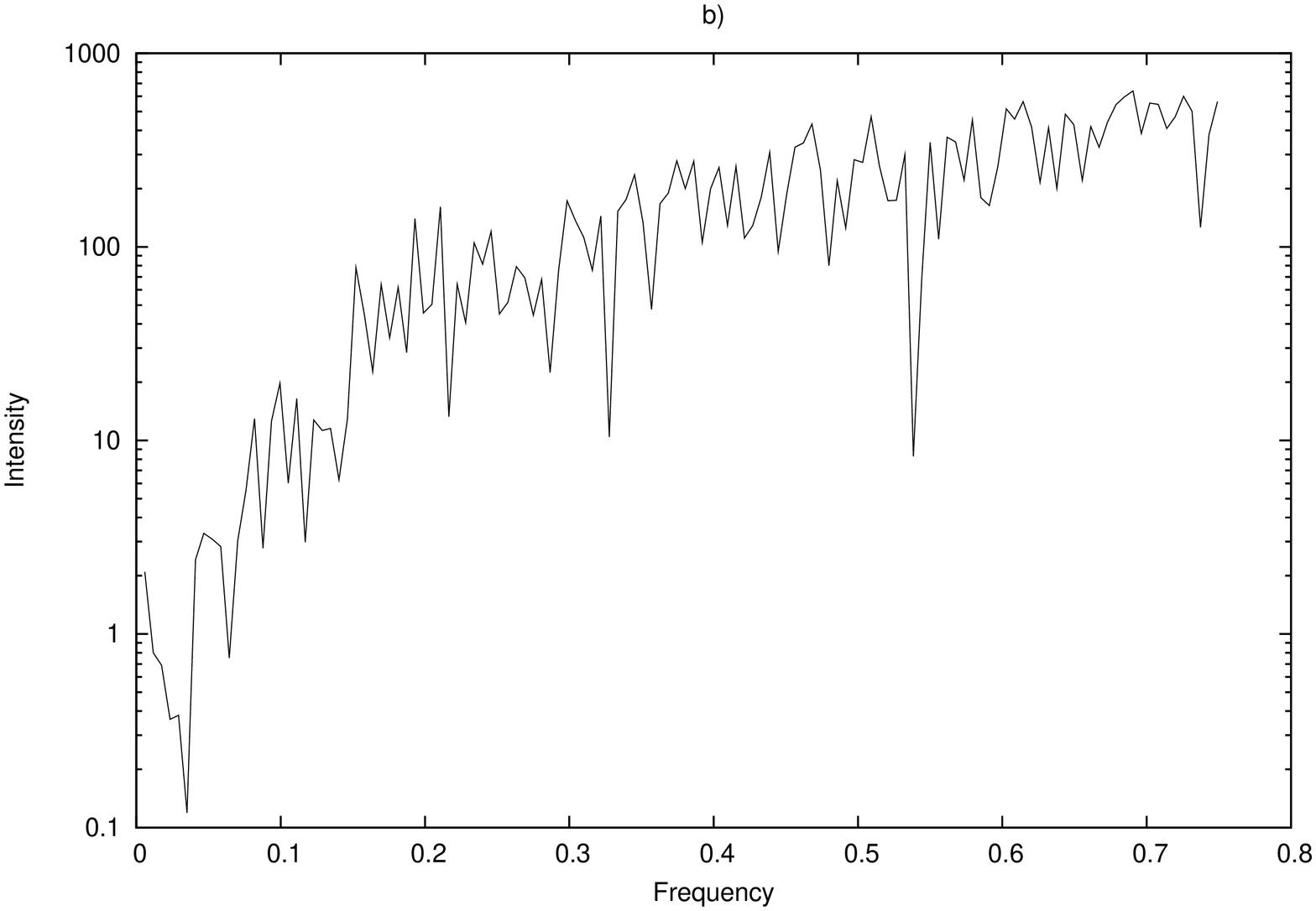}
\caption{Orbital motion of object 546 corresponding to January - September, 2011; a) Time behavior of the semi-major axis and b) Power spectrum of the semi-major axis.} \label{timebehandfreqaviorobj54601}
\end{center}
\end{figure}

\newpage
\begin{figure}[h!]
\begin{center}
	\includegraphics[width=7cm,height=8cm, angle=270]{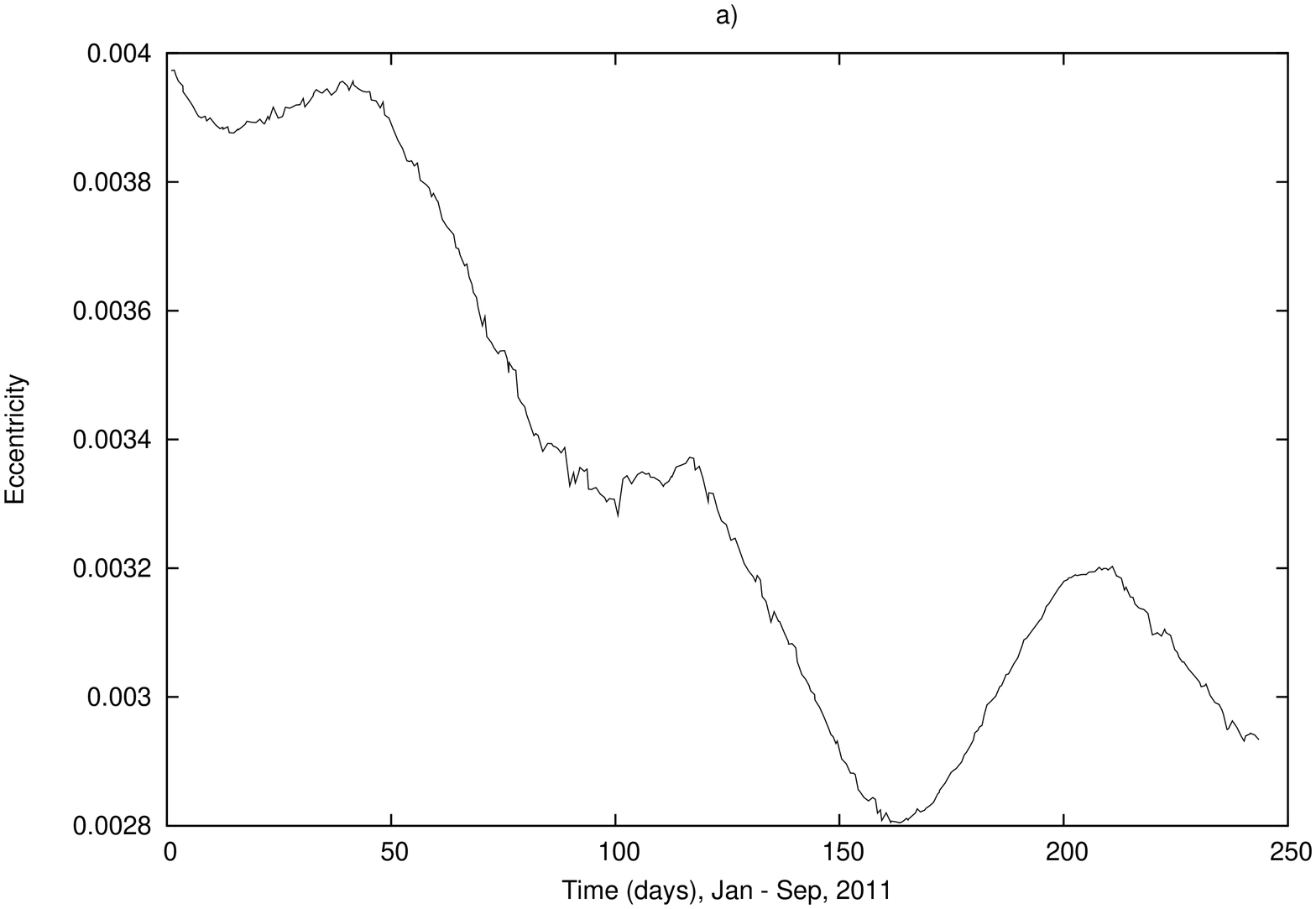} \quad
	\includegraphics[width=7cm,height=7.5cm, angle=270]{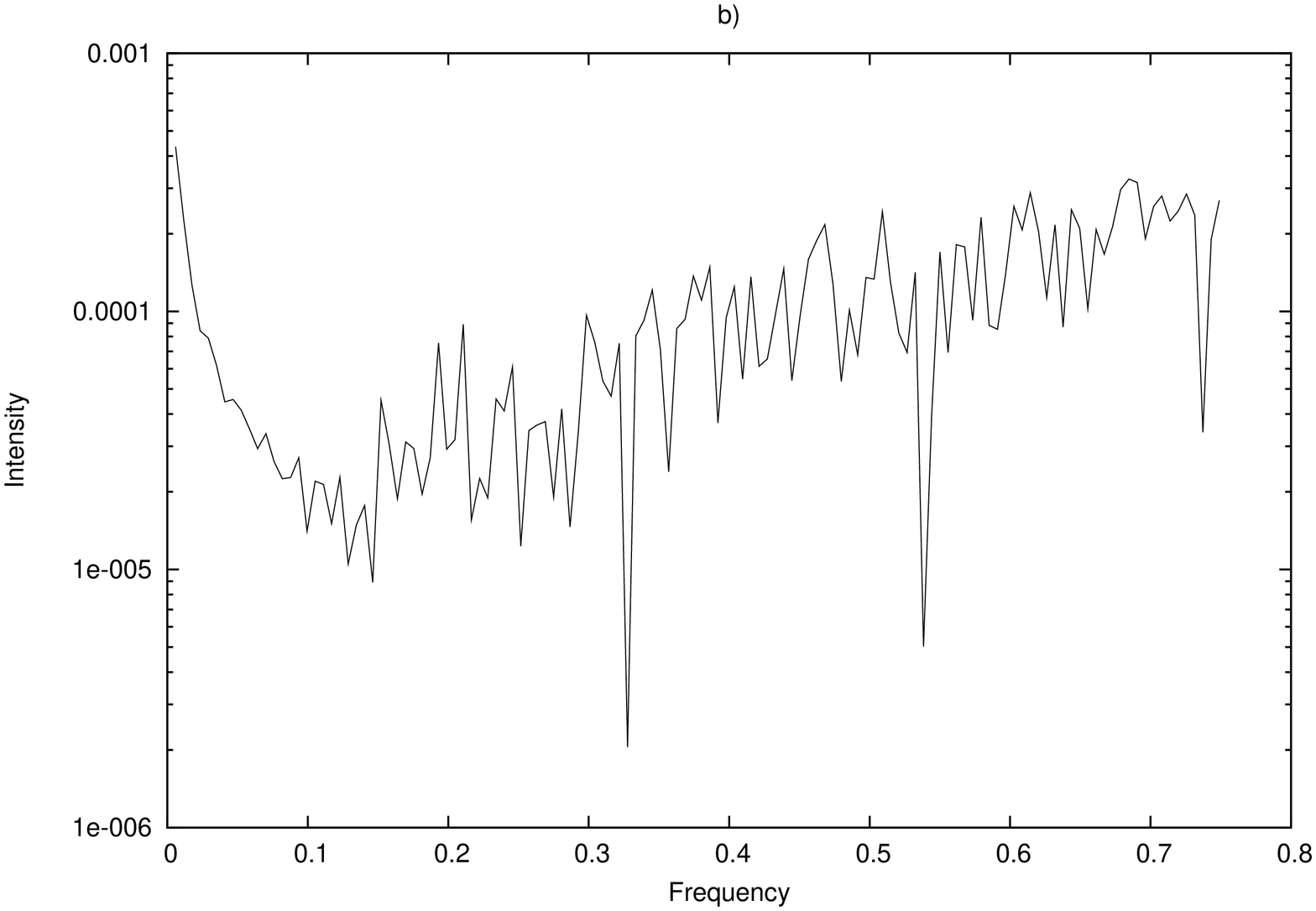}
\caption{Orbital motion of object 546 corresponding to January - September, 2011; a) Time behavior of the eccentricity and b) Power spectrum of the eccentricity.} \label{timebehandfreqaviorobj54602}
\end{center}
\end{figure}

\begin{figure}[h!]
\begin{center}
	\includegraphics[width=7cm,height=8cm, angle=270]{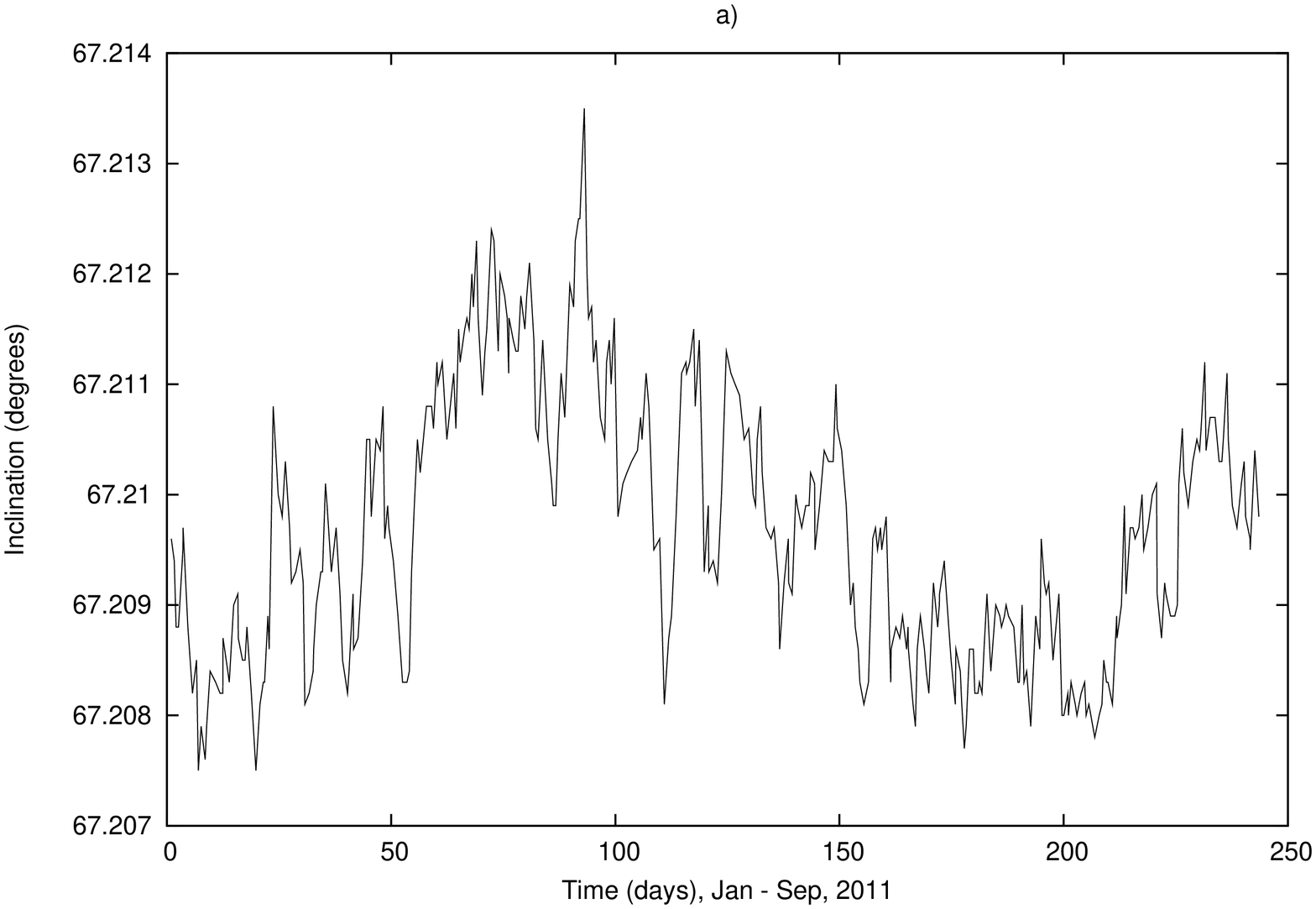} \quad
	\includegraphics[width=7cm,height=7.5cm, angle=270]{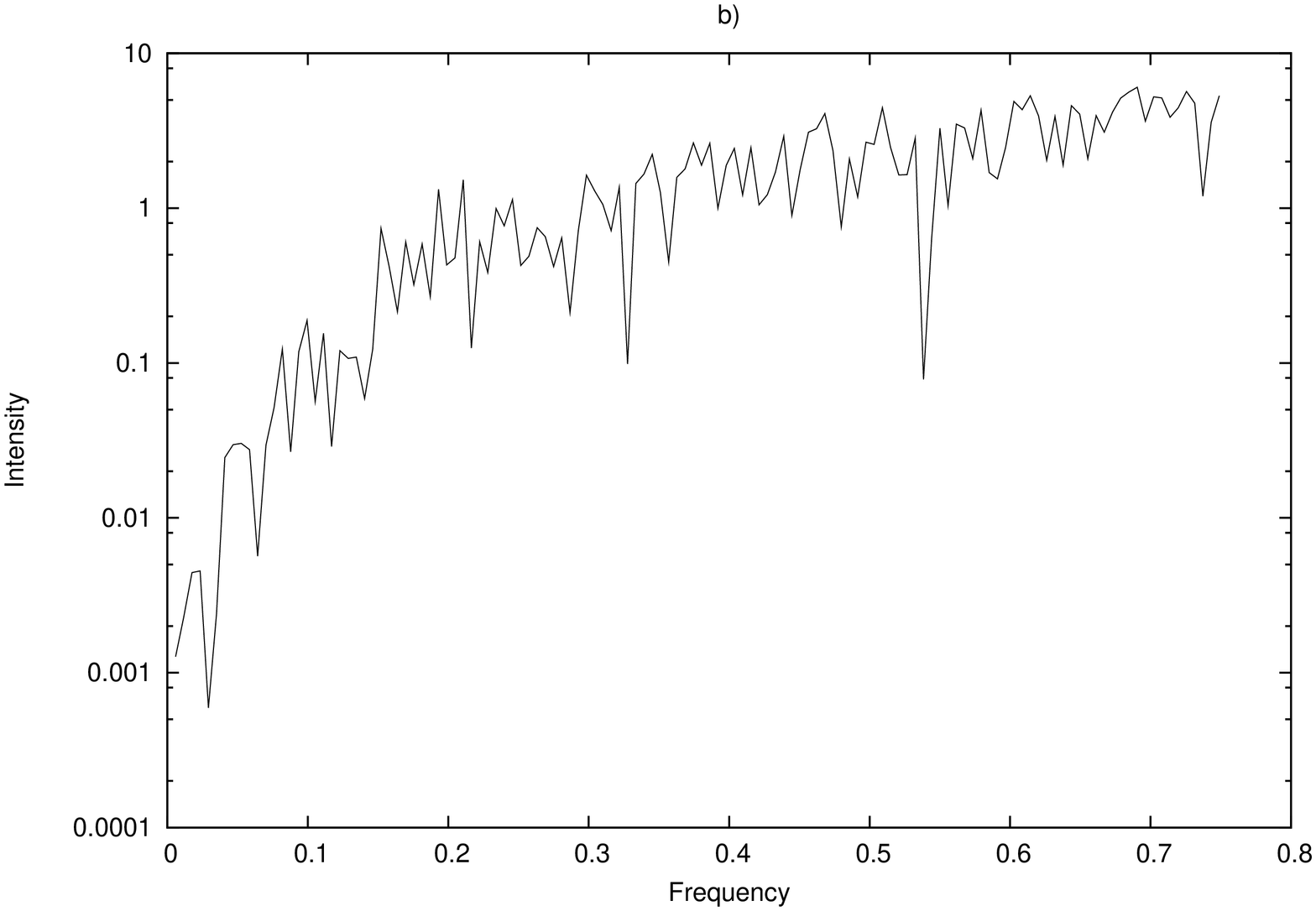}
\caption{Orbital motion of object 546 corresponding to January - September, 2011; a) Time behavior of the inclination and b) Power spectrum of the inclination.} \label{timebehandfreqaviorobj54603}
\end{center}
\end{figure}

\newpage
\begin{figure}[h!]
\begin{center}
	\includegraphics[width=7cm,height=8cm, angle=270]{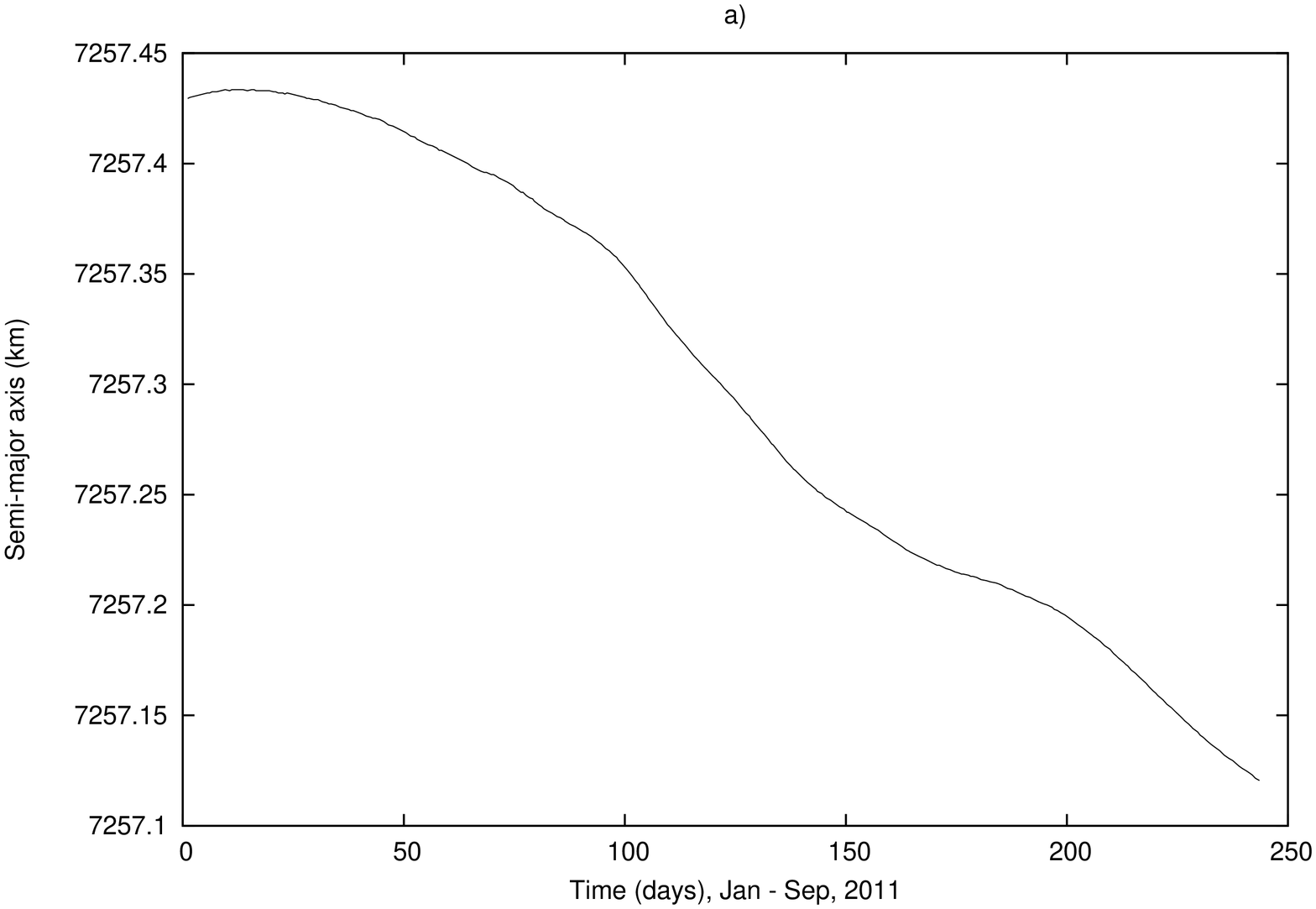} \quad
	\includegraphics[width=7cm,height=7.5cm, angle=270]{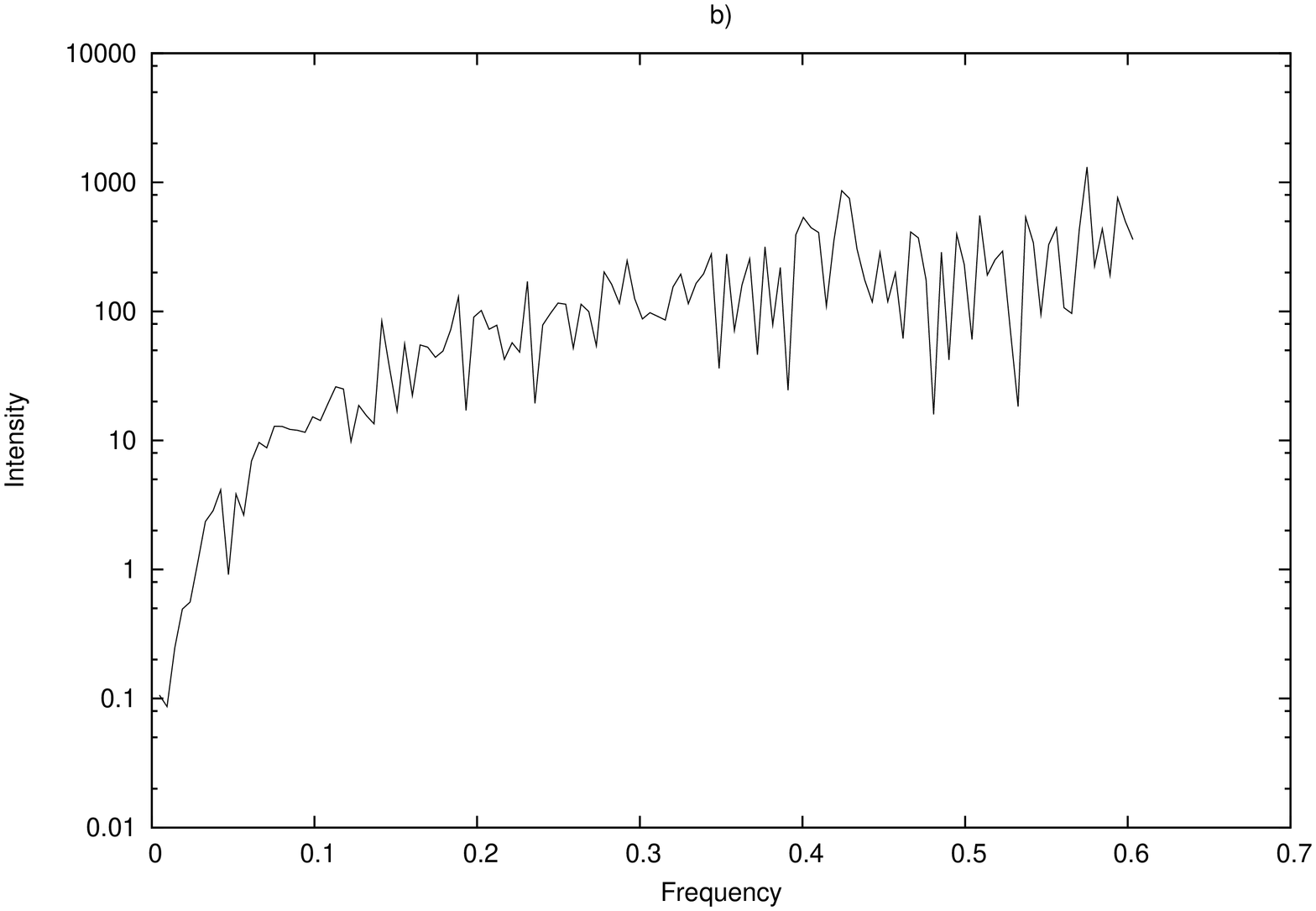}
\caption{Orbital motion of object 2986 corresponding to January - September, 2011; a) Time behavior of the semi-major axis and b) Power spectrum of the semi-major axis.} \label{timebehandfreqaviorobj298601}
\end{center}
\end{figure}

\begin{figure}[h!]
\begin{center}
	\includegraphics[width=7cm,height=8cm, angle=270]{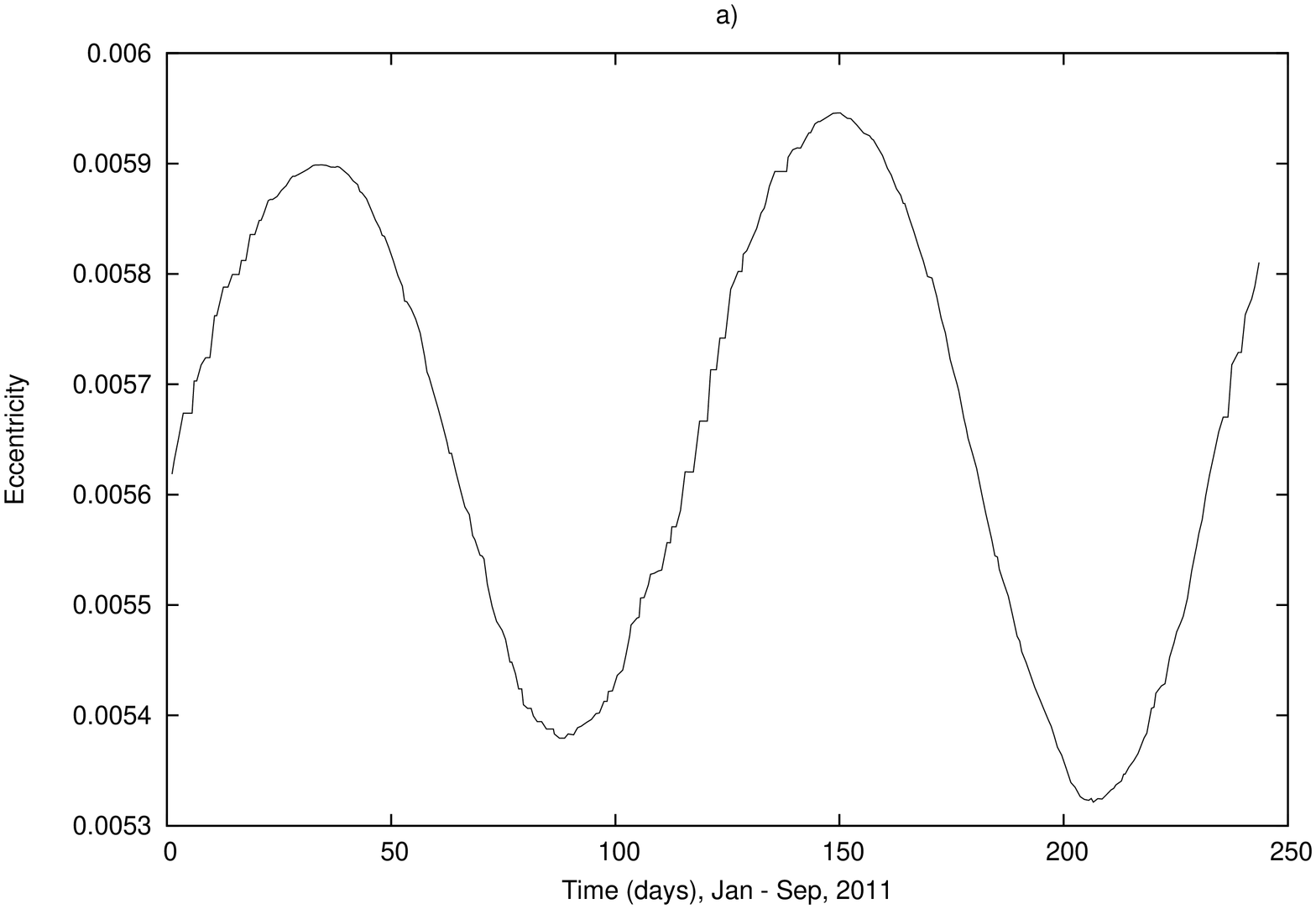} \quad
	\includegraphics[width=7cm,height=7.5cm, angle=270]{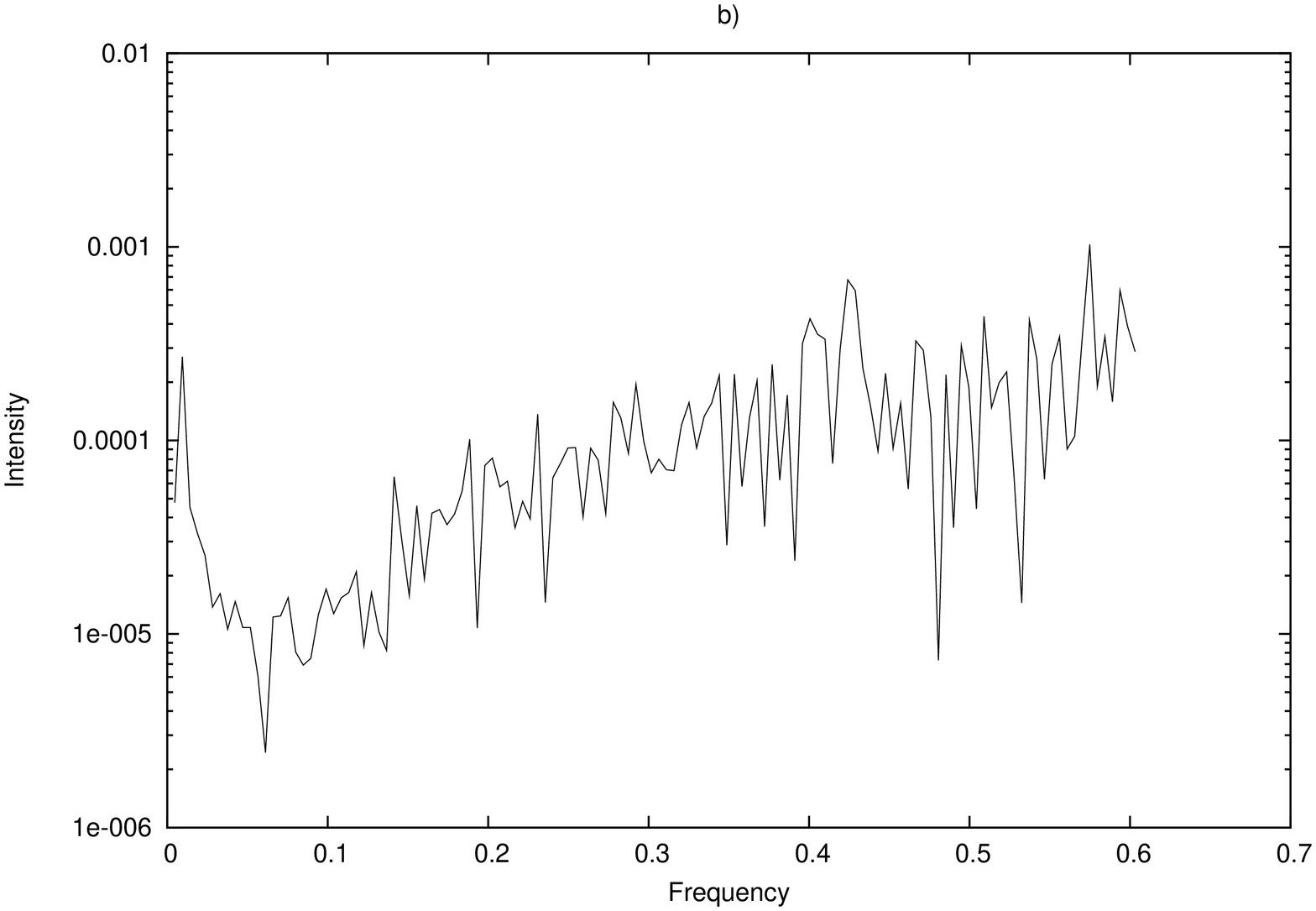}
\caption{Orbital motion of object 2986 corresponding to January - September, 2011; a) Time behavior of the eccentricity and b) Power spectrum of the eccentricity.} \label{timebehandfreqaviorobj298602}
\end{center}
\end{figure}

\newpage
\begin{figure}[h!]
\begin{center}
	\includegraphics[width=7cm,height=8cm, angle=270]{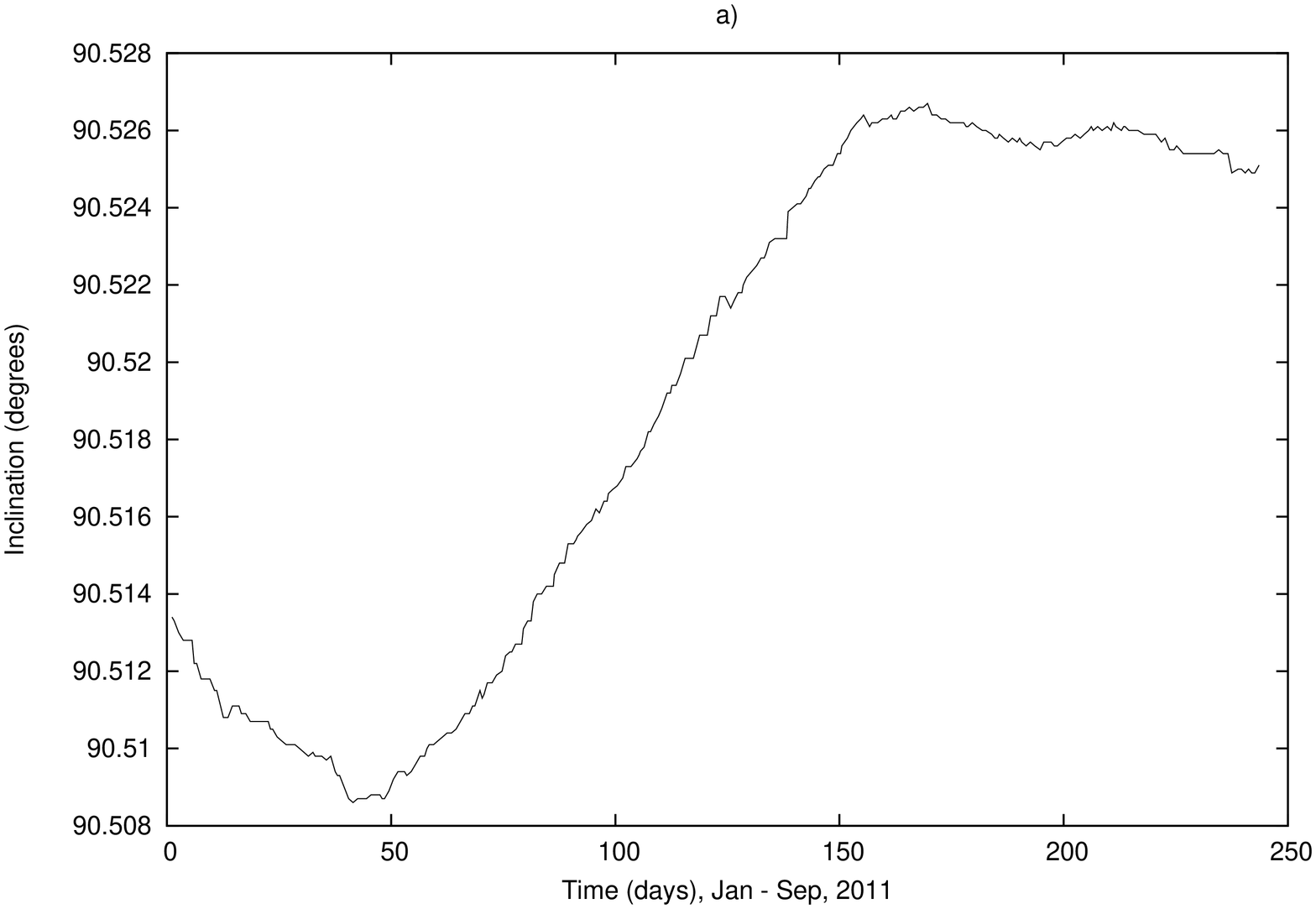} \quad
	\includegraphics[width=7cm,height=7.5cm, angle=270]{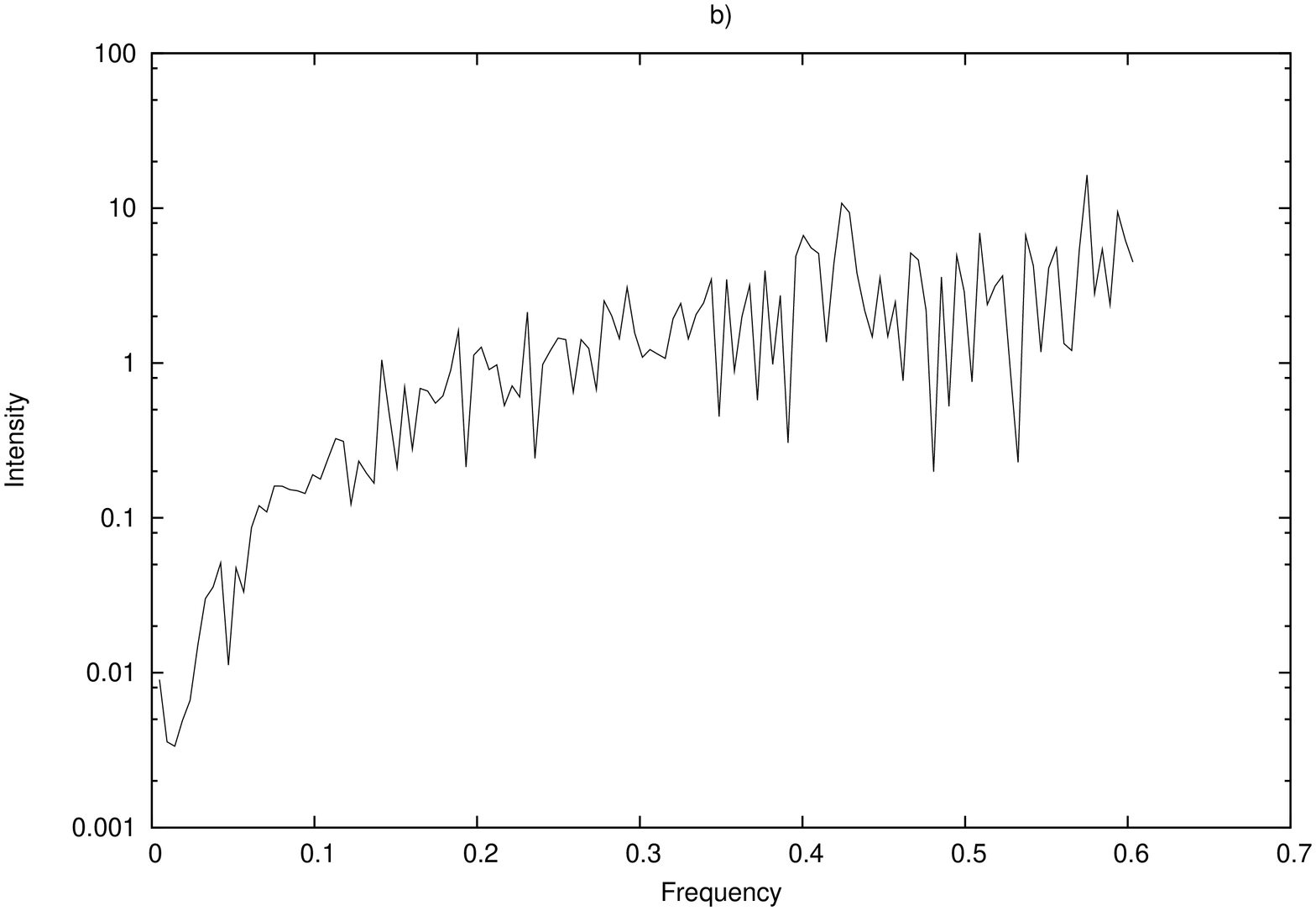}
\caption{Orbital motion of object 2986 corresponding to January - September, 2011; a) Time behavior of the inclination and b) Power spectrum of the inclination.} \label{timebehandfreqaviorobj298603}
\end{center}
\end{figure}

\begin{figure}[h!]
\begin{center}
	\includegraphics[width=7cm,height=8cm, angle=270]{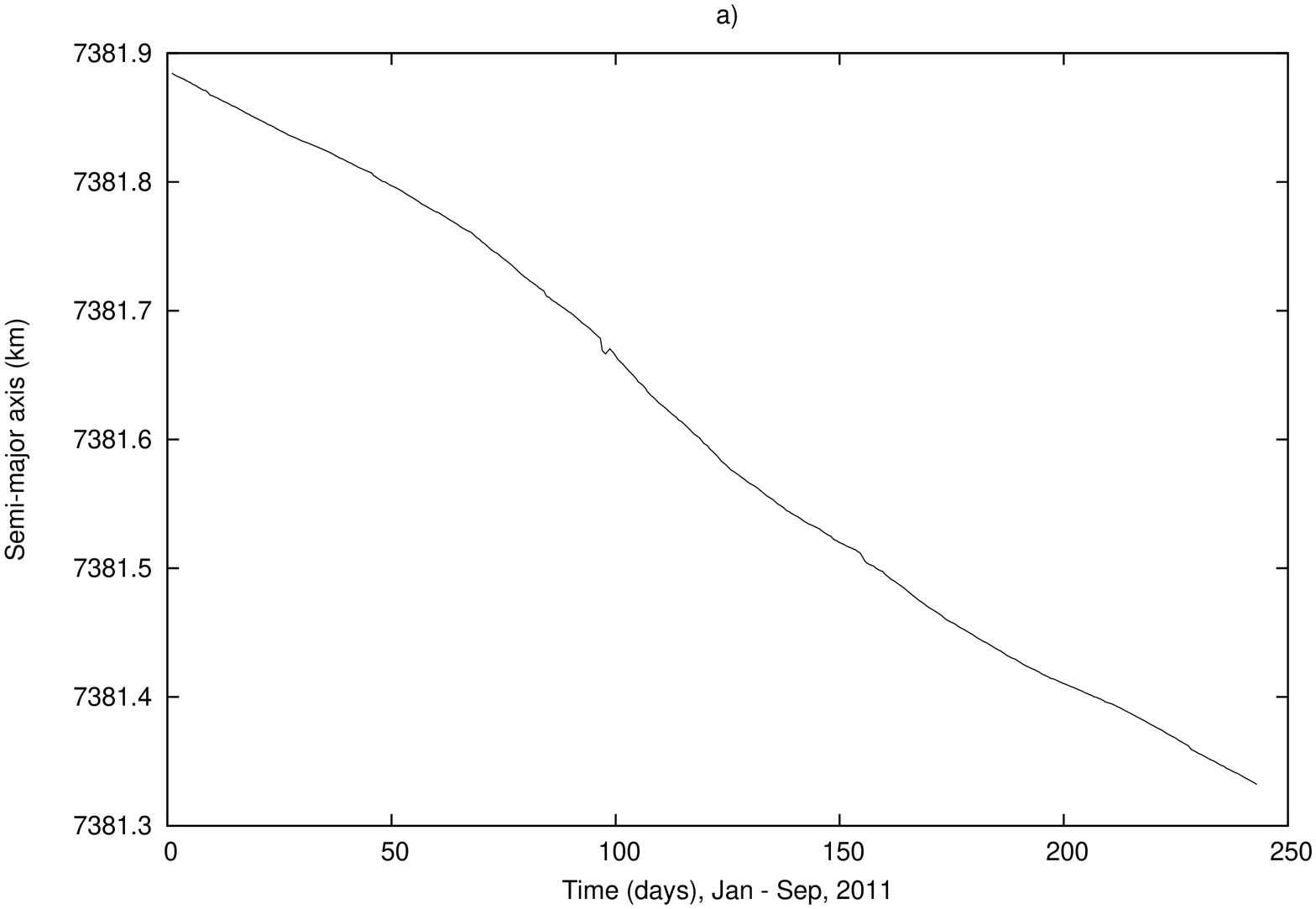} \quad
	\includegraphics[width=7cm,height=7.5cm, angle=270]{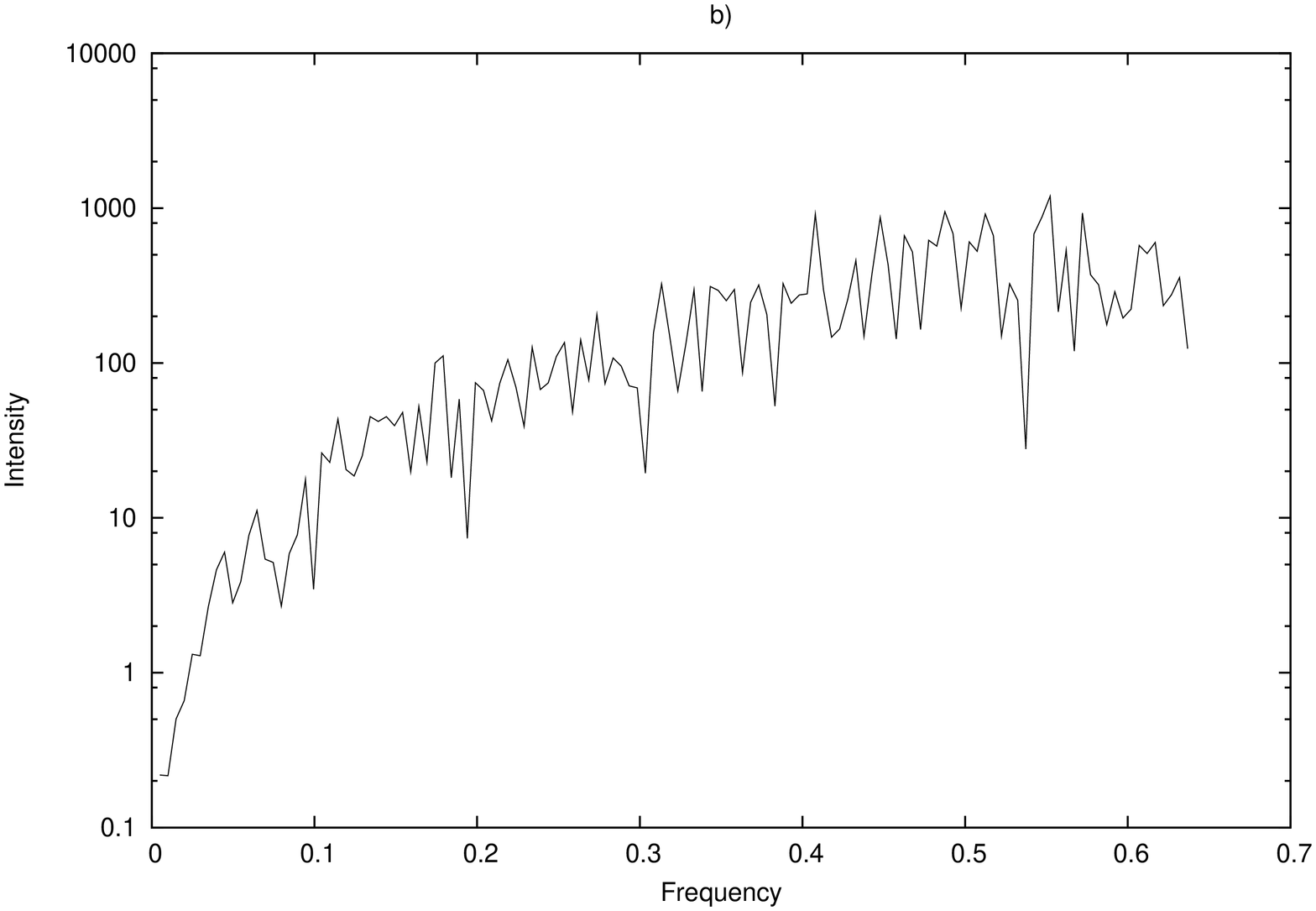}
\caption{Orbital motion of object 4855 corresponding to January - September, 2011; a) Time behavior of the semi-major axis and b) Power spectrum of the semi-major axis.} \label{timebehandfreqaviorobj485501}
\end{center}
\end{figure}

\newpage
\begin{figure}[h!]
\begin{center}
	\includegraphics[width=7cm,height=8cm, angle=270]{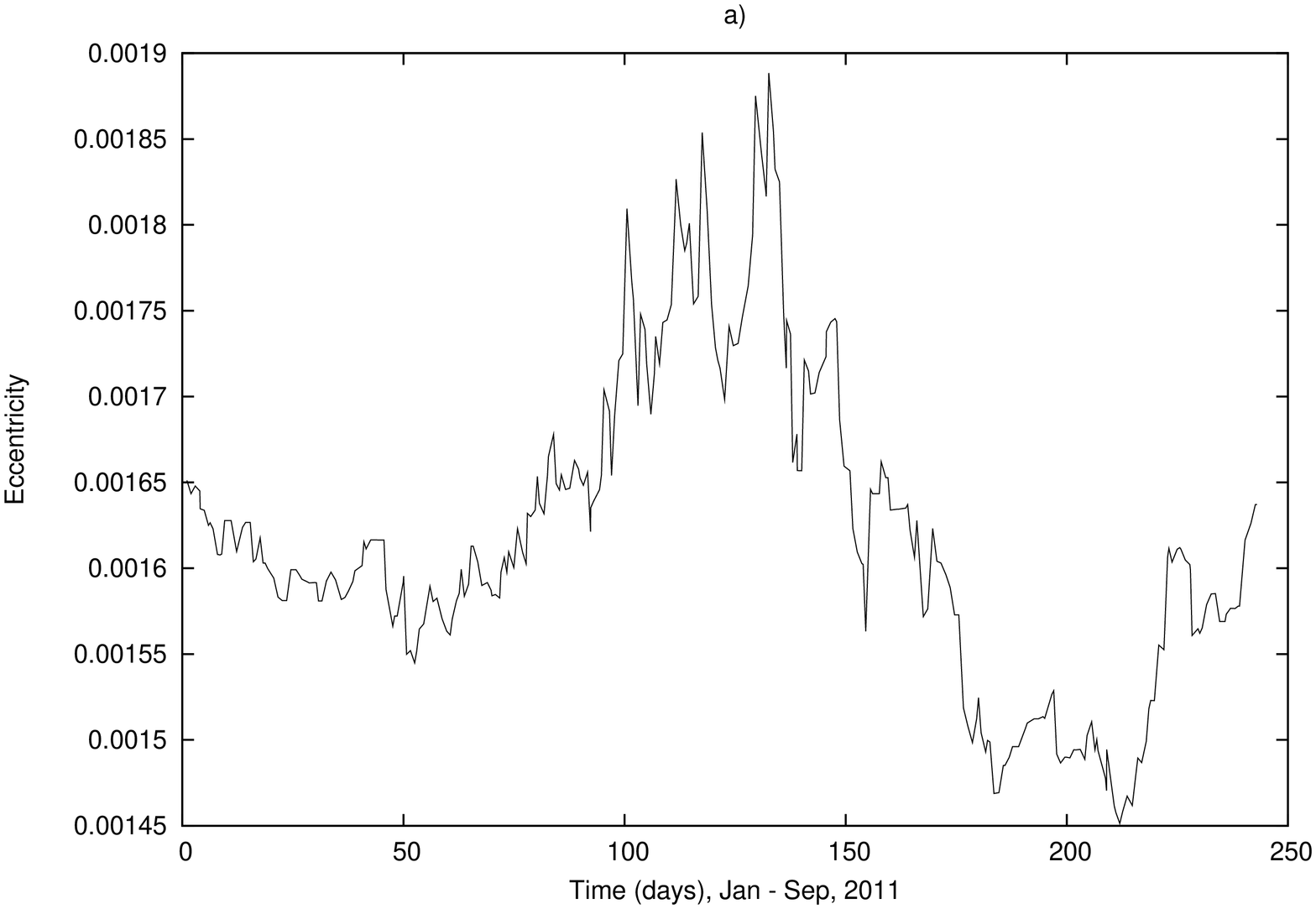} \quad
	\includegraphics[width=7cm,height=7.5cm, angle=270]{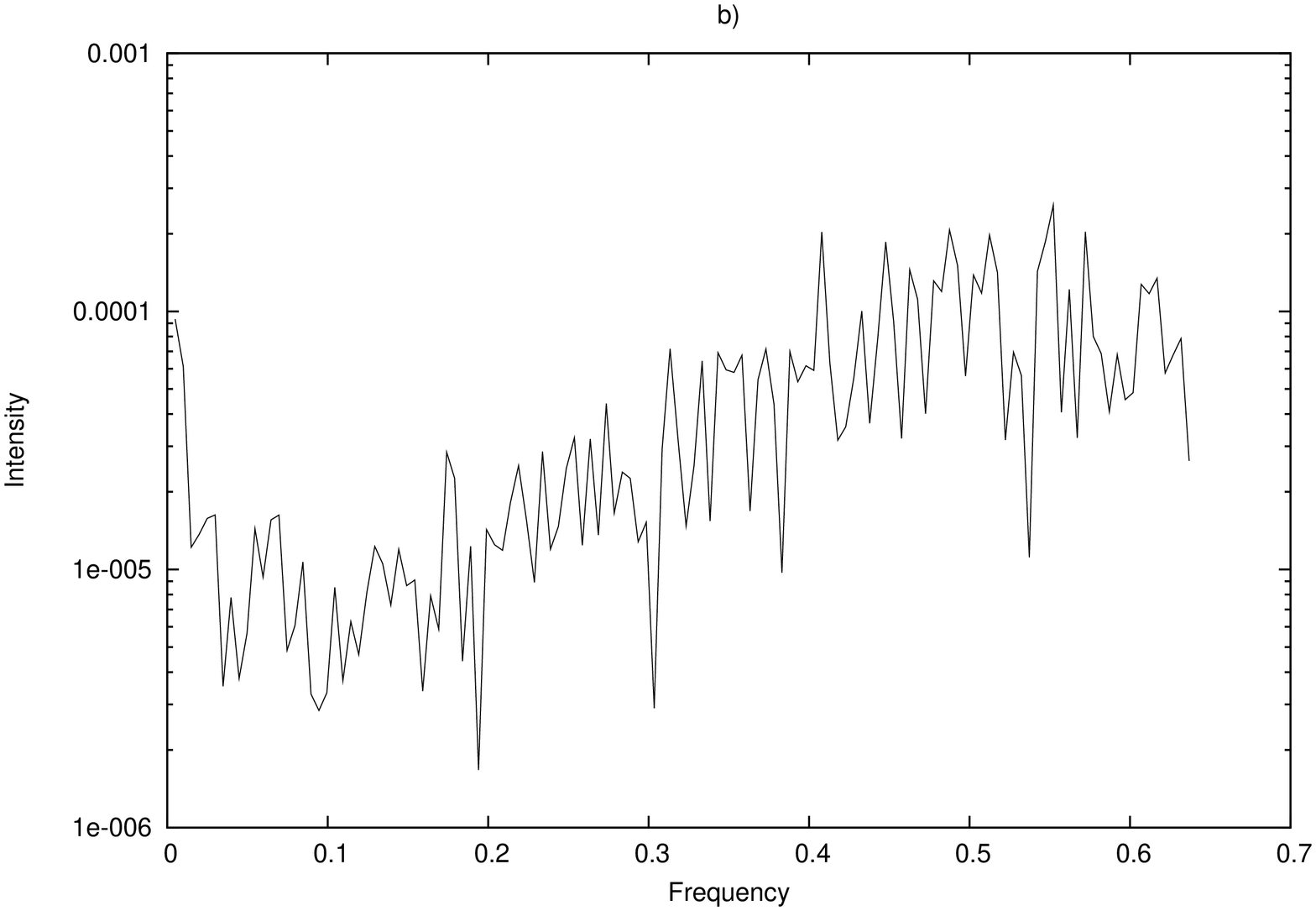}
\caption{Orbital motion of object 4855 corresponding to January - September, 2011; a) Time behavior of the eccentricity and b) Power spectrum of the eccentricity.} \label{timebehandfreqaviorobj485502}
\end{center}
\end{figure}

\begin{figure}[h!]
\begin{center}
	\includegraphics[width=7cm,height=8cm, angle=270]{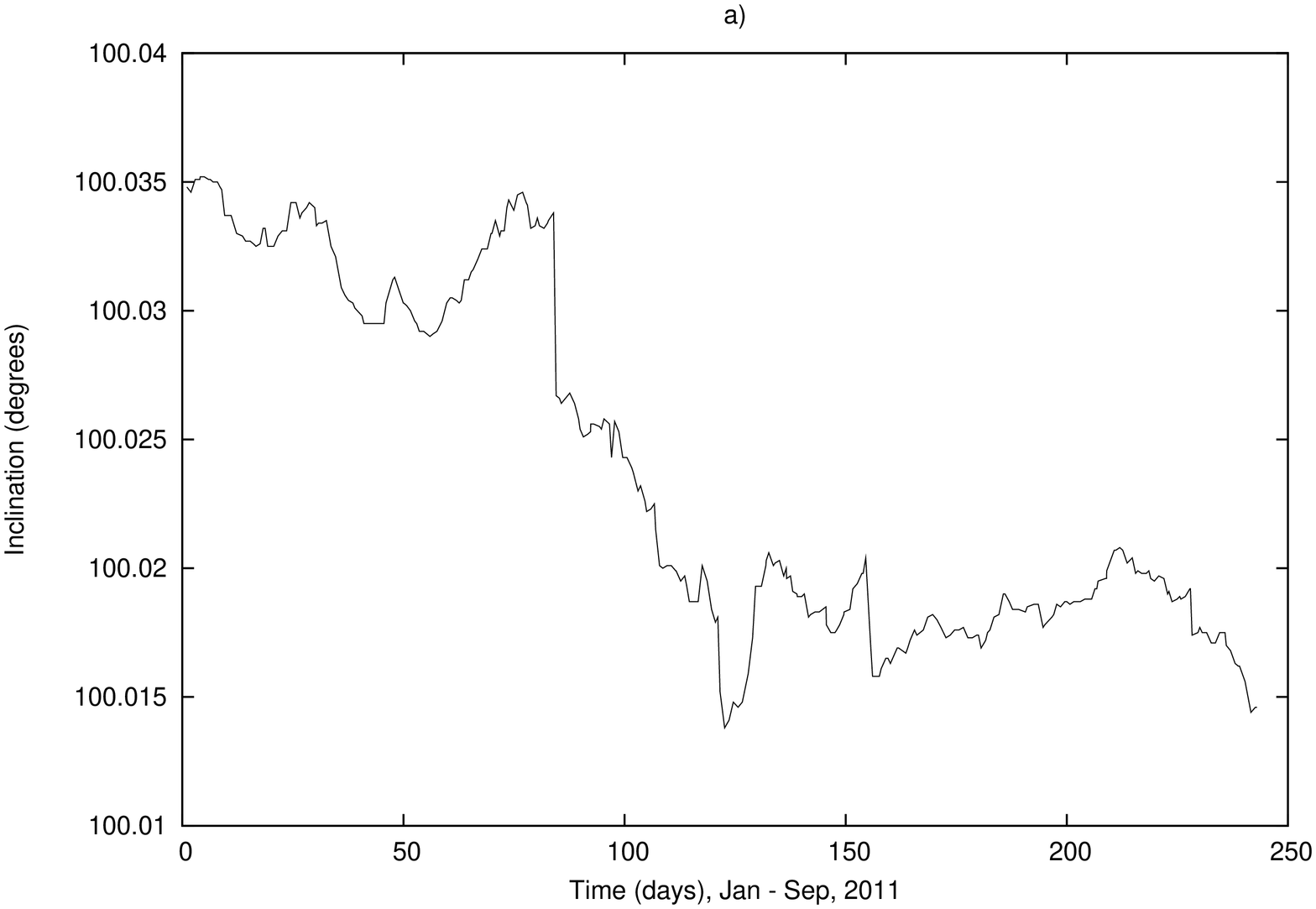} \quad
	\includegraphics[width=7cm,height=7.5cm, angle=270]{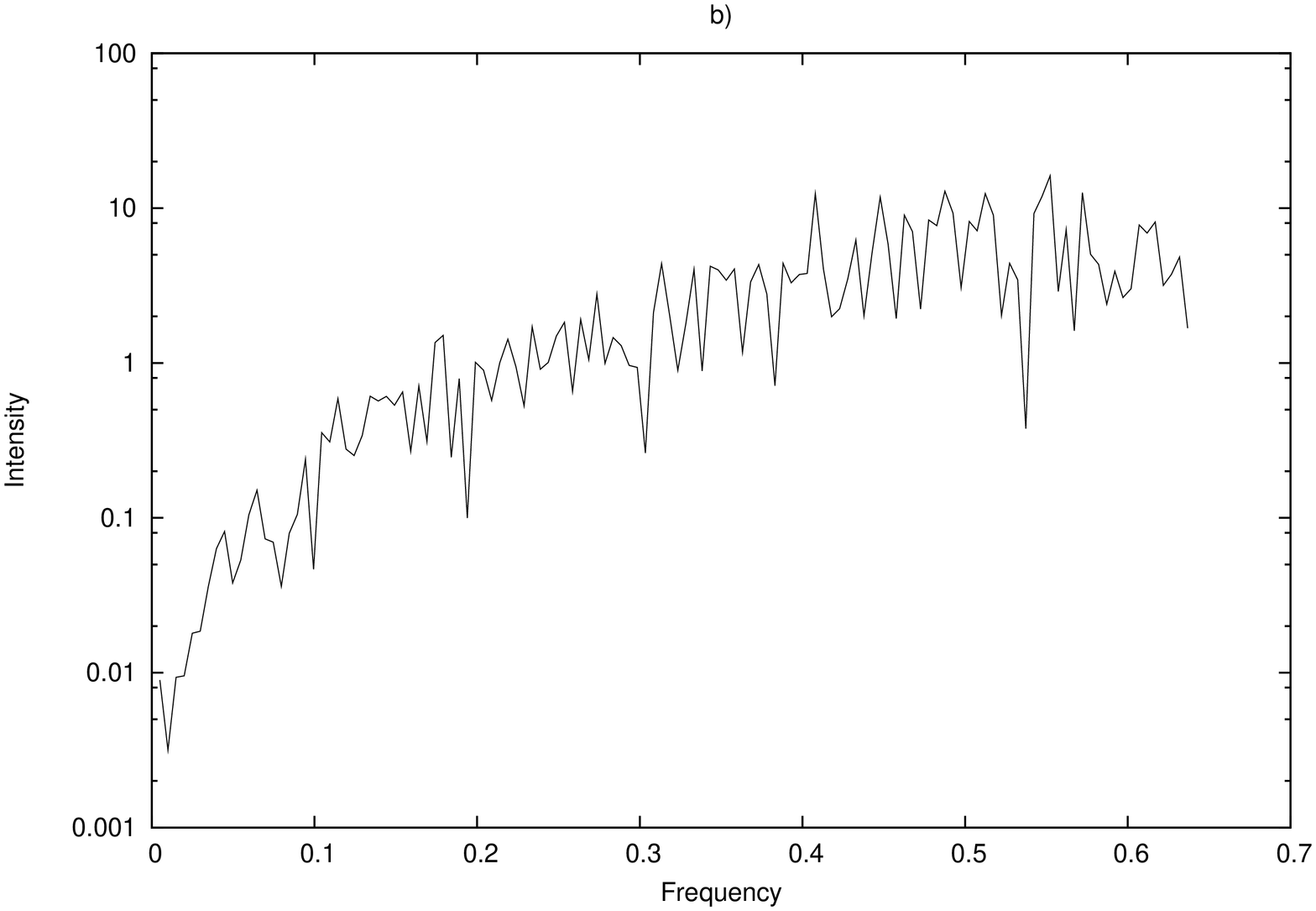}
\caption{Orbital motion of object 4855 corresponding to January - September, 2011; a) Time behavior of the inclination and b) Power spectrum of the inclination.} \label{timebehandfreqaviorobj485503}
\end{center}
\end{figure}

\vspace{0.3cm}
Observing the time behavior of the orbital elements of the objects 325, 546, 2986 and 4855 in Figs. \ref{timebehandfreqaviorobj32501} to \ref{timebehandfreqaviorobj485503}, one can verify possible regular and irregular motions in the trajectories of these space debris. The time behavior of the inclination of the orbital motions of the objects 325, 546 and 2986 show irregularities. Analyzing object 325, note that in 200 days, a fast increase in the inclination occurs but, this variation is about $0.01^{o}$ and it may be related with some disturbance added to the motion.

The power spectrum of the orbital elements are not discrete and they have a big number of frequency components in Figs. \ref{timebehandfreqaviorobj32501} to \ref{timebehandfreqaviorobj485503}. Usually, in the studies of the dynamical systems with the purpose to identify chaos, the frequency analysis is considered for a long time. But, in Low Earth Orbits, a big number of cataloged objects have their motions influenced by different resonant periods and unknown objects can collide with each other providing more unpredictability in their motions. In this case, the frequency analysis can be more appropriated for a short time.

So its also important to observe if the orbital motion is influenced by different resonant angles. In other words, it is verified if the orbital motion is due to overlap resonances.

Tab. \ref{tabcritanglesobjs} shows the resonant angles related to the orbital motions of the objects 325, 546, 2986 and 4855 corresponding to the period January - September, 2011.

  \begin{large}
\begin{table}[h!]
\caption{Resonant angles $\phi_{kmq}$ related to the orbital motions of the objects 325, 546, 2986 and 4855.}
\centering
\large{
\begin{tabular}{c c c c}
\hline \normalsize{Object} & \normalsize{Coefficient $k$} & \normalsize{Coefficient $m$} &  \normalsize{Coefficient $q$}     \\
\hline 325  & 1  & 14  & 0 \\ \hline 325 & 6 & 42 & -3 \\ \hline 325 & 4 & 28 & -2 \\ \hline 325 & 2 & 14 & -1 \\ \hline
325  & 7  & 42  & -4 \\ \hline 325 & 5 & 28 & -3 \\ \hline 325 & 3 & 14 & -2 \\ \hline 325 & 8 & 42 & -5 \\ \hline
325  & 6  & 28  & -4 \\ \hline 325 & 4 & 14 & -3 \\ \hline 325 & 7 & 28 & -5 \\ \hline 546 & 3 & 43 & 0 \\ \hline
546  & 4  & 43  & -1 \\ \hline 546 & 5 & 43 & -2 \\ \hline 546 & 6 & 43 & -3 \\ \hline 546 & 7 & 43 & -4 \\ \hline
546  & 8  & 43  & -5 \\ \hline 2986 & -1 & 28 & 3 \\ \hline 2986 & -1 & 42 & 4 \\ \hline 2986 & 0 & 14 & 1 \\ \hline
4855  & 7  & 41  & -4 \\ \hline 4855 & 8 & 41 & -5 \\ \hline
\end{tabular}}
\label{tabcritanglesobjs}
\end{table}
\end{large}

Observing Tab. \ref{tabcritanglesobjs}, one can verify that the orbital dynamics of the objects 2986 and 4855 are influenced by some resonant angles, while for the objects 325 and 546, several resonant angles influence their orbits simultaneously.

If the commensurability between the orbital motions of the object and the Planet is defined by the parameter $\alpha$ and by the condition $\alpha=(k+q)/m$, one can say that the exact 14:1 resonance is defined by the condition $\alpha=1/14$. This way, analyzing Tab. \ref{tabcritanglesobjs}, it is verified that the motions of the objects 325 and 2986 are influenced by the exact 14:1 resonance, while the objects 546 and 4855 are influenced by resonant angles in the neighborhood of the exact resonance.

Figures \ref{timebehresperobj325} to \ref{timebehresperobj4855} show the time behavior of the resonant period corresponding to the resonant angles presented in Tab. \ref{tabcritanglesobjs}.

\newpage
\begin{figure}[h!]
\centering
 \includegraphics[scale=1.3, width=9cm, angle=270]{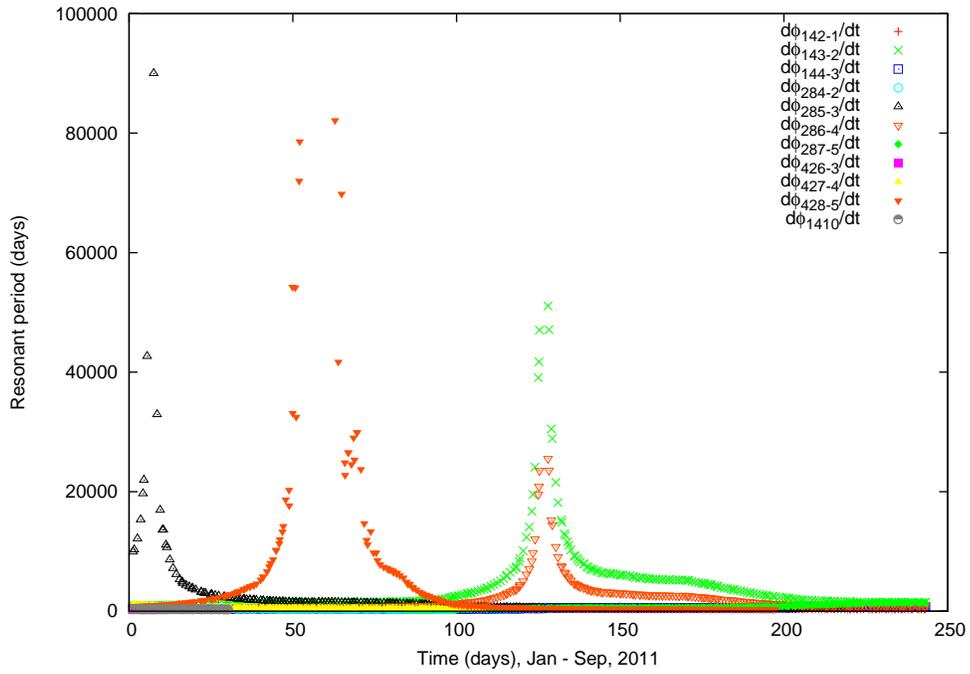}
\caption{Time behavior of the resonant period corresponding to the orbital motion of object 325.}
\label{timebehresperobj325}
\end{figure}

\vspace{0.3cm}
\begin{figure}[h!]
\centering
 \includegraphics[scale=1.3, width=9cm, angle=270]{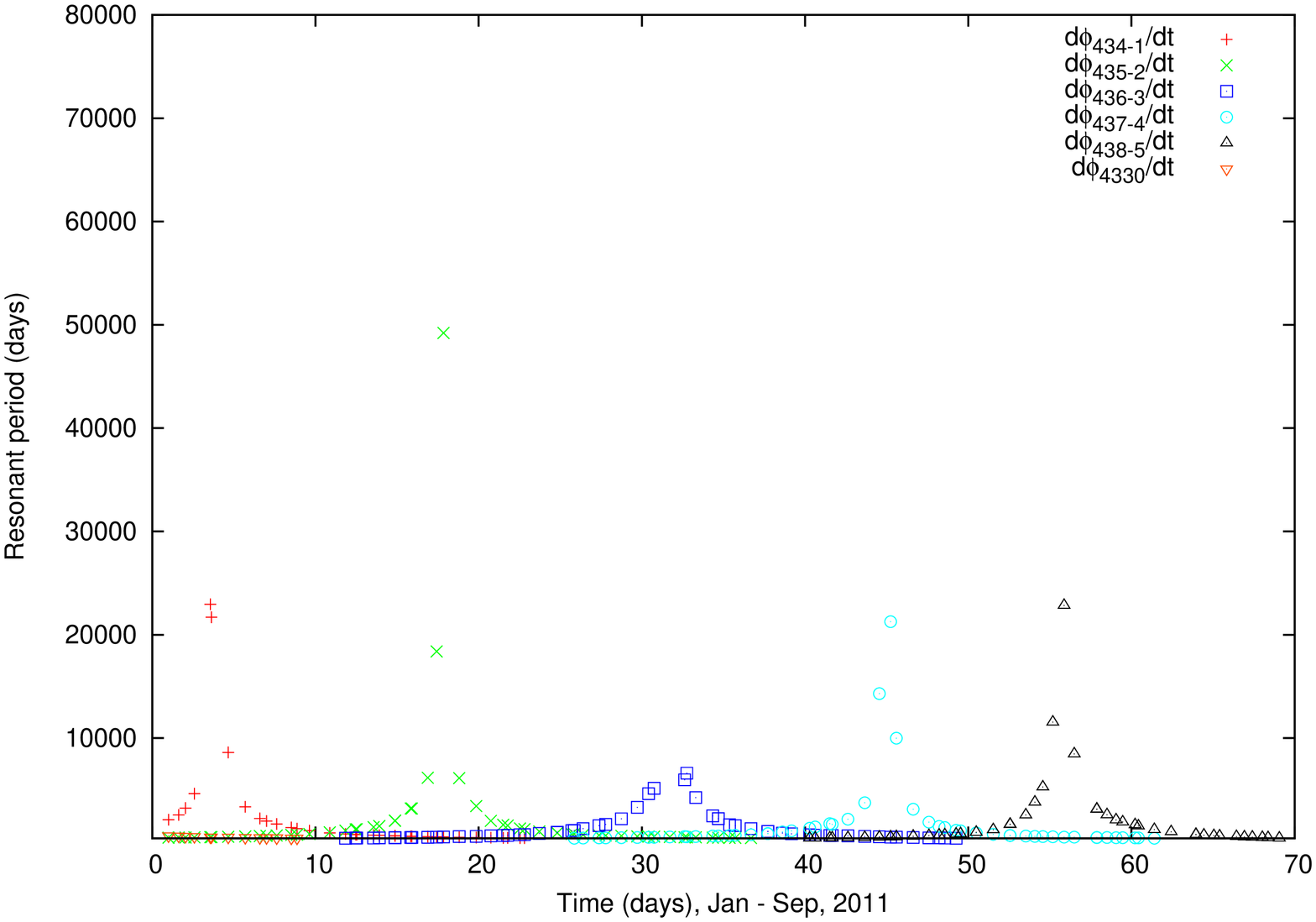}
\caption{Time behavior of the resonant period corresponding to the orbital motion of object 546.}
\label{timebehresperobj546}
\end{figure}

\newpage

\begin{figure}[h!]
\centering
 \includegraphics[scale=1.3, width=9cm, angle=270]{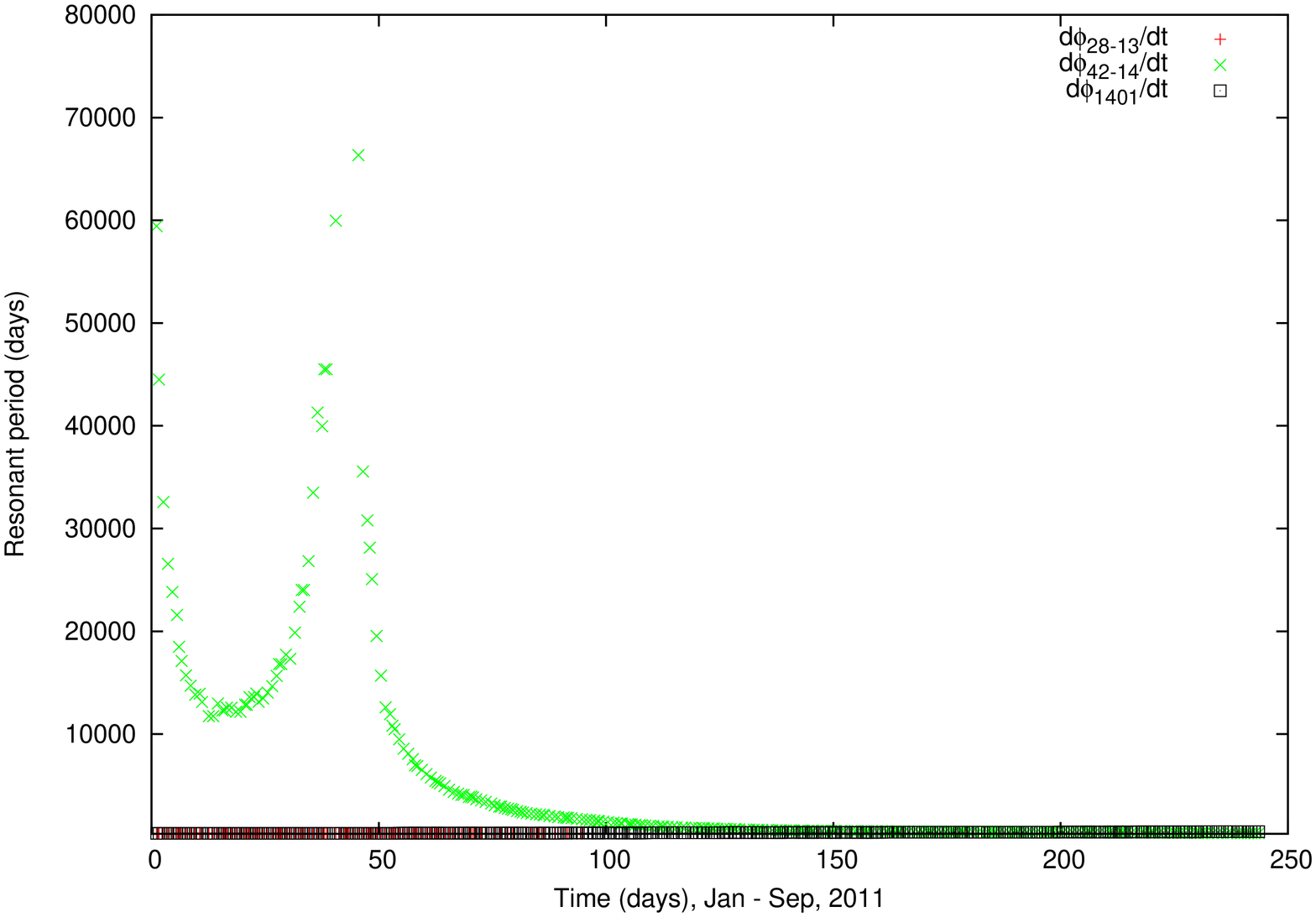}
\caption{Time behavior of the resonant period corresponding to the orbital motion of object 2986.}
\label{timebehresperobj2986}
\end{figure}

\begin{figure}[h!]
\centering
 \includegraphics[scale=1.3, width=9cm, angle=270]{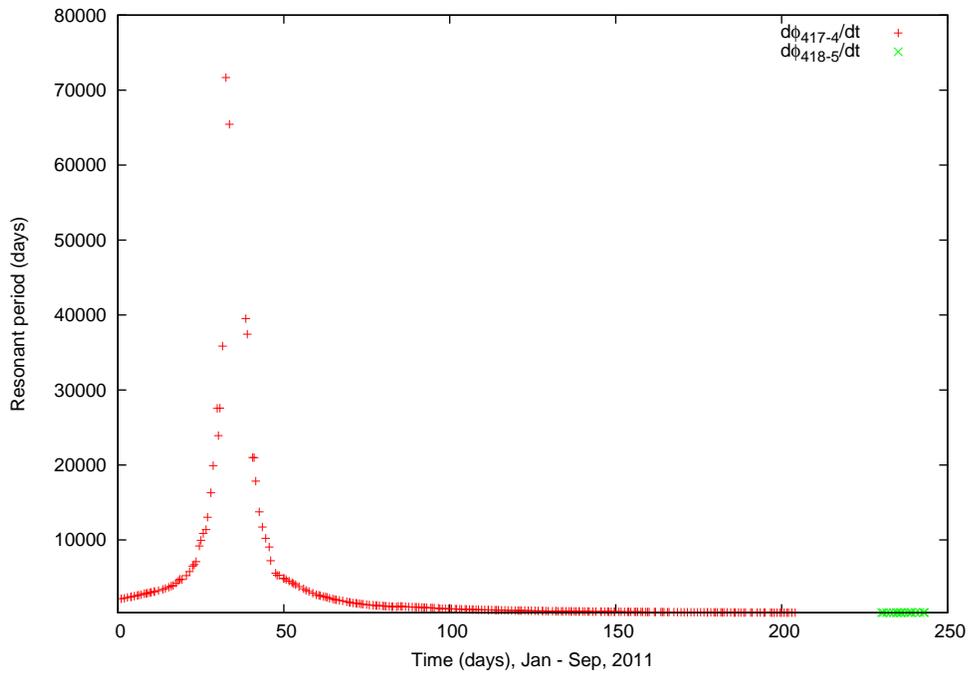}
\caption{Time behavior of the resonant period corresponding to the orbital motion of object 4855.}
\label{timebehresperobj4855}
\end{figure}

\newpage
Analyzing the time behavior of the resonant period in Figs. \ref{timebehresperobj325} to \ref{timebehresperobj4855} it is verified that the resonant angles remain in deep resonance for a few days. The orbital motion of object 546 has resonant angles which satisfies the established criterium $Pres > 300$ days for 70 days and consequently the orbital dynamics of this object may be losing the influence of the 14:1 resonance. Figures \ref{timebehresperobj2986} and \ref{timebehresperobj4855} corresponding to the objects 2986 and 4855 show the resonant period for values greater than 10000 days in the first fifty days, after that decreasing for values smaller than 500 days in the resonant period and it shows a tendency to not remain in deep resonance. The orbital motion of object 325 is more influenced by the 14:1 resonance and this fact can be observed by the time behavior of the $\dot{\phi}_{kmq}$, Fig. \ref{timebehphidotobj325}.

To continue the analysis about the irregular orbital motions, the time behavior of the $\dot{\phi}_{kmq}$ is studied verifying if different resonant angles describe the orbital dynamics of these objects at the same moment. This fact can indicate irregular orbits requiring a full system with all resonant angles as described in Tab. \ref{tabcritanglesobjs}. Figures \ref{timebehphidotobj325} to \ref{timebehphidotobj4855} show the time behavior of the $\dot{\phi}_{kmq}$ corresponding to the resonant angles presented in Tab. \ref{tabcritanglesobjs}.

\vspace*{0.4cm}
\begin{figure}[h!]
\centering
 \includegraphics[scale=1.3, width=9cm, angle=270]{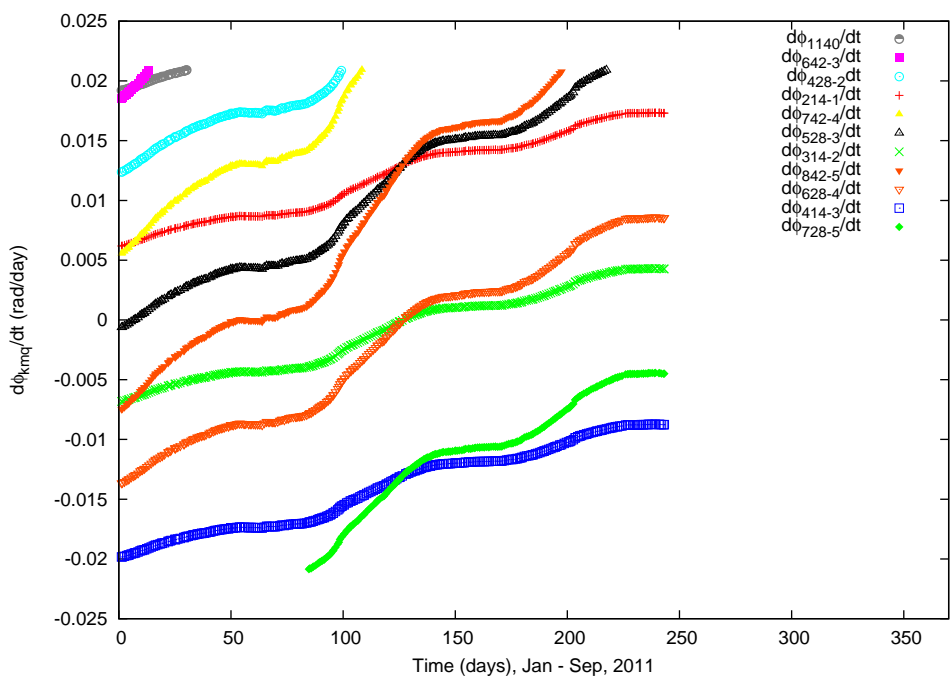}
\caption{Time behavior of $\dot{\phi}_{kmq}$ corresponding to the orbital motion of object 325.}
\label{timebehphidotobj325}
\end{figure}

\newpage
\begin{figure}[h!]
\centering
 \includegraphics[scale=1.3, width=9cm, angle=270]{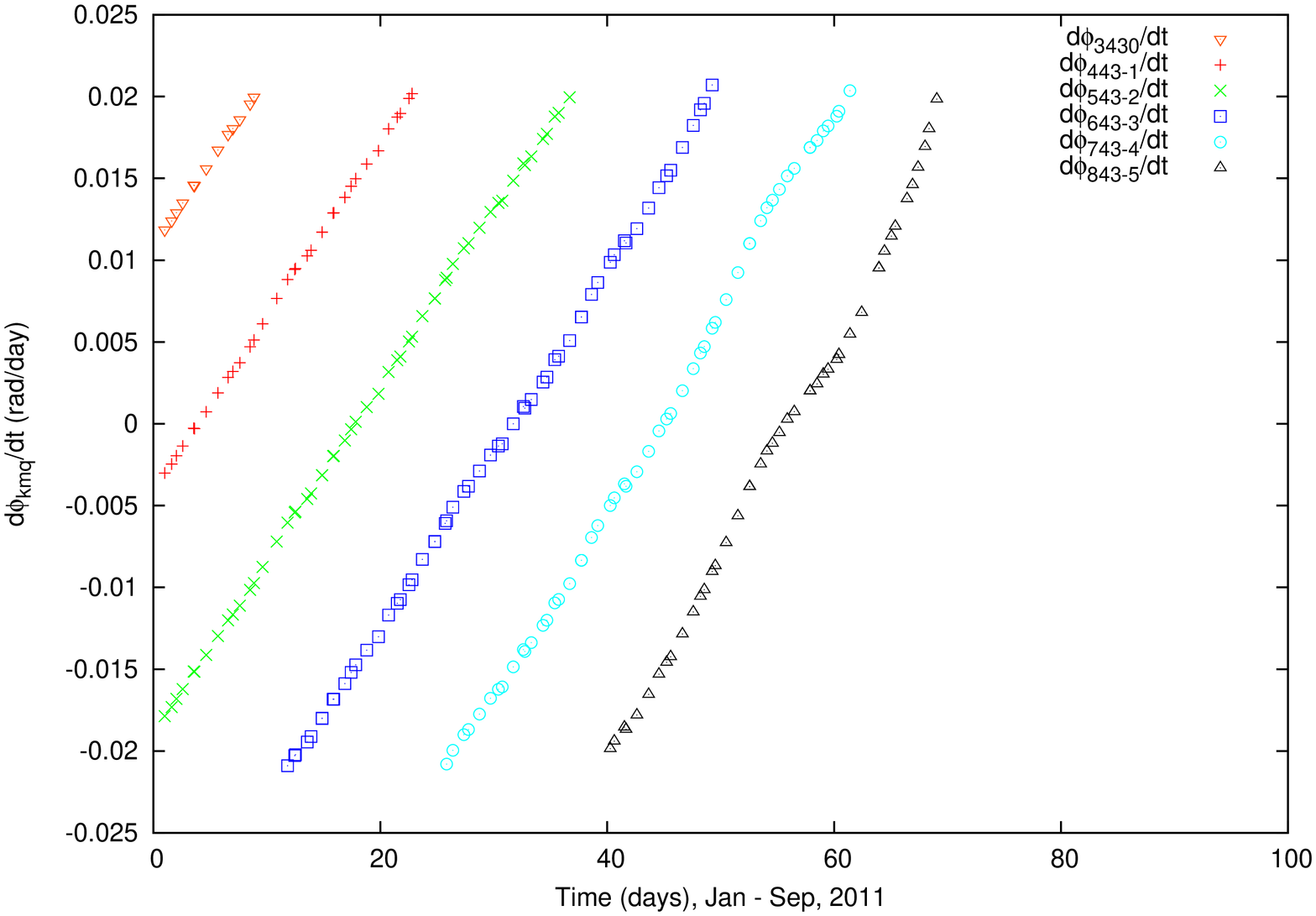}
\caption{Time behavior of $\dot{\phi}_{kmq}$ corresponding to the orbital motion of object 546.}
\label{timebehphidotobj546}
\end{figure}

\vspace*{0.3cm}
\begin{figure}[h!]
\centering
 \includegraphics[scale=1.3, width=9cm, angle=270]{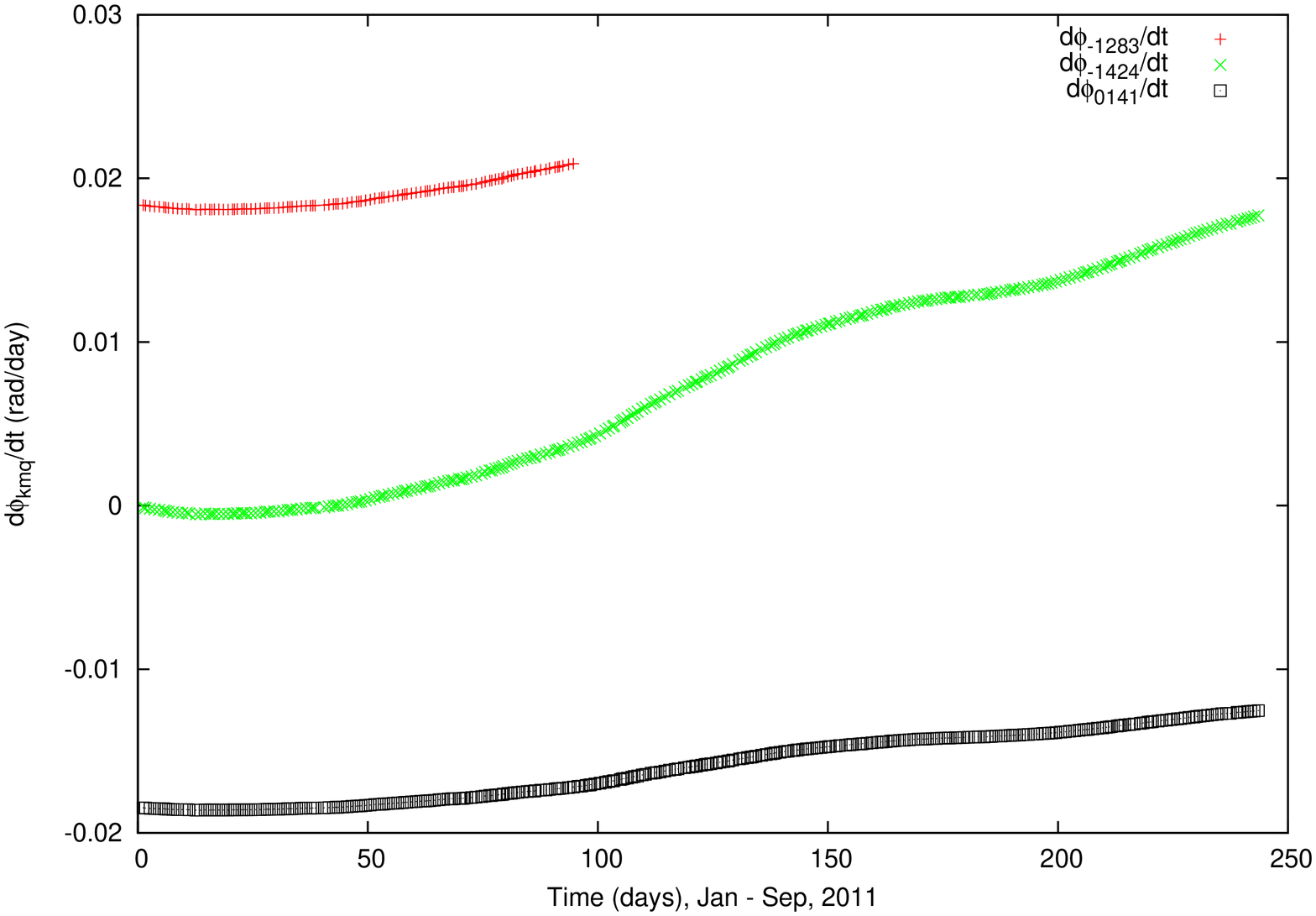}
\caption{Time behavior of $\dot{\phi}_{kmq}$ corresponding to the orbital motion of object 2986.}
\label{timebehphidotobj2986}
\end{figure}

\newpage
\begin{figure}[h!]
\centering
 \includegraphics[scale=1.3, width=9cm, angle=270]{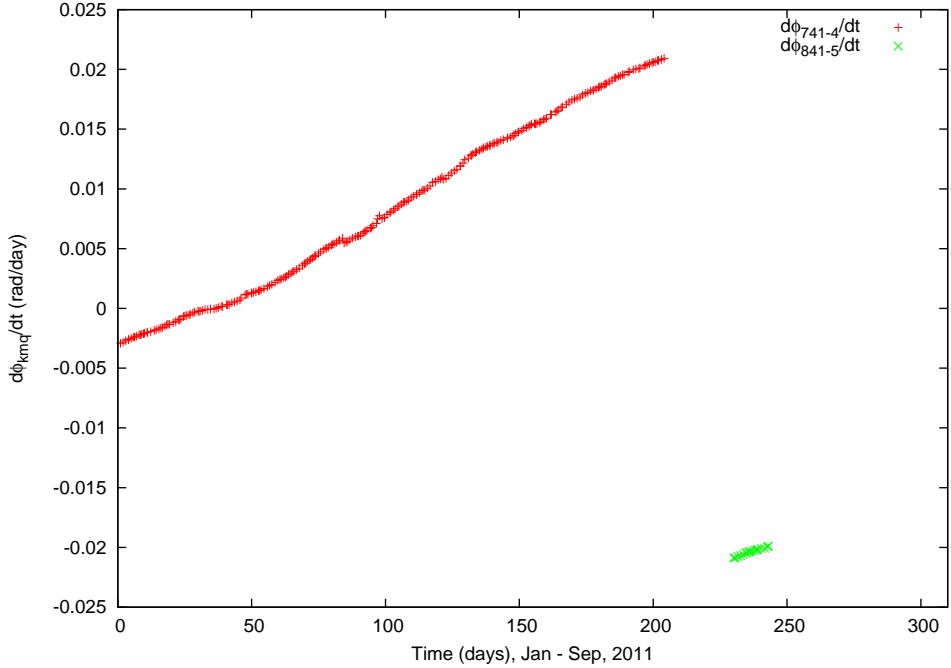}
\caption{Time behavior of $\dot{\phi}_{kmq}$ corresponding to the orbital motion of object 4855.}
\label{timebehphidotobj4855}
\end{figure}

Analyzing the time behavior of the resonant angles $\dot{\phi}_{kmq}$ in Figs. \ref{timebehphidotobj325} to \ref{timebehphidotobj4855} it is possible to observe irregular orbital motion for object 325. In the beginning, the orbital motion of object 546 is governed by distinct resonant angles but, after seventy days, the influence of resonant effects decreases or disappears. Objects 325 and 2986 have their orbital motions influenced by the exact 14:1 resonance and they need a full system with different resonant angles which compose their motions. Otherwise, the orbital motion of object 4855 is defined for resonant angles separately, which can be defined as a regular orbit.

The results and discussions show the complexity in the orbital dynamics of these objects caused by the resonance effects. Furthermore, the increasing number of space debris and collisions between them can cause big problems for artificial satellites missions.

\section*{\normalsize 4. Conclusions}

In this work, the orbital dynamics of synchronous space debris are studied. From the TLE data of the NORAD, objects orbiting the Earth in deep resonance are investigated.

Analyzing the cataloged objects satisfying the established criterium of the resonant period greater than 300 days, four space debris (325, 546, 2986 and 4855) are studied observing the irregular characteristics in their orbits. The time behavior of the orbital elements and the respective frequency analysis are used to verify the orbits of these objects.

The time behavior of the resonant period of the objects show large amplitudes, and consequently, the resonant angles alternate between deep and short resonance.

Several resonant angles influence, simultaneously, the orbital motions of objects 325 and 2986. The time behavior of the inclination and the frequency analysis confirm the irregularities in their orbits. The orbital dynamics of object 546 is firstly influenced by several resonant angles in the neighborhood of the 14:1 resonance but, after some days, the regular characteristic in its orbit seems to dominate. The time behavior of the $\dot{\phi}_{kmq}$ of object 4855 shows resonant angles separately indicating a regular orbit.

The TLE catalog of the objects orbiting the Earth is an important tool to study the orbital motion of the artificial satellites and space debris.

\section*{\small {ACKNOWLEDGEMENTS}}
This work was accomplished with support of the FAPESP under the contract N$^{o}$ 2009/00735-5 and 2006/04997-6, SP-Brazil, and CNPQ
(contracts 300952/2008-2 and 302949/2009-7).

\renewcommand\bibname{\small{References}}


\begin{thebibliography}{20}
\thispagestyle{empty}
\bibitem{osiander} R. Osiander, P. Ostdiek, Introduction to Space Debris, Handbook of Space Engineering, Archeology and Heritage, 2009.
\bibitem{ikeda} H. Ikeda, T. Hanada, T. Yasaka, Searching for lost fragments in GEO, Acta Astronautica 63 (2008) 1312–1317.
\bibitem{spacetrack} Space Track. Archives of the 2-lines elements of NORAD. Available at: <www.space-track.org>, accessed in February-September, 2011.
\bibitem{kuga} H. K. Kuga, Utilização das efemérides "2-lines" do NORAD no modelo orbital do satélite CBERS-1, I Congresso de Dinâmica, Controle e Aplicações, Issue 1 (2002) 925-930.
\bibitem{orlando} V. Orlando, R. V. F. Lopes, H. K. Kuga, Flight dynamics team experience throughout four years of SCD1 in-orbit operations: main issues, improvements and trends, in: Proceedings of XII International Symposium on Space Flight Dynamics, Darmstadt, Germany, ESOC, 1997, pp. 433-437.
\bibitem{space} F. R. Hoots, R. L. Roehrich, Models for Propagation of NORAD Element Sets, Spacetrack Report N$^{o}$. 3, 1980.
\bibitem{changyin} Z. Changyin, Z. Wenxiang, H. Zengyao, W. Hongbo, Progress in space debris research, Chinese Journal of Space
Science 30 (2010) 516-518.
\bibitem{nishida} S. Nishida, S. Kawamoto, Y. Okawa, F. Terui, S. Kitamura, Space debris removal system using a small satellite, Acta Astronautica 65 (2009) 95-102.
\bibitem{mechishnek} M. J. Mechishnek, Overview of the Space Debris Environment, AEROSPACE REPORT NO. TR-95(5231)-3, 1995.
\bibitem{morando} M. B. Morando,  Orbites de Resonance des Satellites de 24h, Bull. Astron. 24 (1963) pp. 47.
\bibitem{blitzer} L. Blitzer, Synchronous and Resonant Satellite Orbits Associated with Equatorial Ellipticity, ARS Journal 32 (1963) 1016-1019.
\bibitem{garfinkel3} B. Garfinkel,  Formal Solution in the Problem of Small Divisors, Astron. Journal 71 (1966) 657-669.
\bibitem{gedeon1} G. S. Gedeon, O. L. Dial,  Along-track Oscillations of a Satellite due to Tesseral Harmonics, AIAA Journal 5 (1967) 593-595.
\bibitem{lane} M. T. Lane,  An Analytical Treatment of Resonance Effects on Satellite Orbits, Celestial Mechanics 42 (1988) 3-38.
\bibitem{jupp} A. Jupp,  A Solution of the Ideal Resonance Problem for the Case of Libration, Astron. Journal 74 (1969) 35-43.
\bibitem{ely} T. A. Ely, K. C. Howell,  Long-term Evolution of Artificial Satellite Orbits due to Resonant Tesseral Harmonics, Journal of the Astronautical Sciences 44 (1996) 167-190.
\bibitem{lima} P. H. C. N. Lima Jr.,  Sistemas Ressonantes a Altas Excentricidades no Movimento de Satélites Artificiais, Tese de Doutorado, Instituto Tecnológico de Aeronáutica, 1998.
\bibitem{grosso} P. R. Grosso,   Movimento Orbital de um Satélite Artificial em Ressonância 2:1, Tese de Mestrado, Instituto Tecnológico de Aeronáutica, 1989.
\bibitem{diogo} D. M. Sanckez, T. Yokoyama, P. I. O. Brasil, R. R. Cordeiro, Some Initial Conditions for Disposed Satellites of
the Systems GPS and Galileo Constellations, Mathematical Problems in Engineering, 2009.
\bibitem{ferreira} L. D. D. Ferreira and R. Vilhena de Moraes, "GPS Satellites Orbits: Resonance", Mathematical Problems in Engineering, Vol. 2009.
\bibitem{sampaio} J. C. Sampaio, R. Vilhena de Moraes, S. S. Fernandes, Artificial Satellites Dynamics: Resonant Effects, in: Proceedings of the 22nd International Symposium on Space Flight Dynamics, São José dos Campos, 2011.
\bibitem{rossi} A. Rossi, Resonant dynamics of Medium Earth Orbits: space debris issues, Celestial Mechanics and Dynamical Astronomy 100 (2008) 267–286.
\bibitem{neto} A. G. S. Neto, Estudo de Órbitas Ressonantes no Movimento de Satélites Artificiais, Tese de Mestrado, ITA, 2006.
\bibitem{deleflie} F. Deleflie, A. Rossi, C. Portman, G. Métris, F. Barlier, Semi-analytical investigations of the long term evolution of the eccentricity of Galileo and GPS-like orbits, Advances in Space Research 47, Issue 5 (2011) 811-821.
\bibitem{anselmo} L. Anselmo, C. Pardini, Dynamical evolution of high area-to-mass ratio debris released into GPS orbits, Advances in Space Research 43, Issue 10 (2009) 1491-1508.
\bibitem{chao} C. C. Chao, R. A. Gick, Long-term evolution of navigation satellite orbits: GPS/GLONASS/GALILEO, Advances in Space Research 34, Issue 5 (2004) 1221-1226.
\bibitem{sampaiompe} J. C. Sampaio, R. Vilhena de Moraes, S. S. Fernandes, The Orbital Dynamics of Synchronous Satellites: Irregular Motions in the 2:1 Resonance, Mathematical Problems in Engineering, 2012.
\bibitem{anselmo2} L. Anselmo, C. Pardini, Orbital evolution of the first upper stages used for the new European and Chinese navigation systems, Acta Astronautica 68 (2011) 2066 - 2079.
\bibitem{sampaiocobem} J. C. Sampaio, R. Vilhena de Moraes, S. S. Fernandes, Terrestrial Artificial Satellites Dynamics: Resonant Effects, in: Proceedings of the 21st International Congress of Mechanical Engineering, Natal, 2011.
\bibitem{justina} J. Golebiwska, E. Wnuk, I. Wytrzyszczak, Space debris observation and evolution predictions, Software Algorithms Document (SAD), ESA PECS Project N$^{o}$ 98088, Astronomical Observatory of the Adam Mickiewicz University, Poznan, 2010.
\bibitem{powell} G. E. Powell, I. C. Percival, Spectral Entropy method for distinguishing regular and irregular motions in Hamiltonian systems, J. Phys. A, Math. Gen. 12 (1979) 2053-2071.
\bibitem{laskar} J. Laskar, Frequency analysis for multi-dimensional systems. Global dynamics and diffusion, Physica D 67 (1993) 257-281.
\bibitem{milos} M. Sidlichovsky, D. Nesvorny, Frequency Modified Fourier Transform and its Application to Asteroids, Celestial Mechanics and Dynamical Astronomy 65 (1997) 137-148.
\bibitem{callegarijr} N. Callegari Jr., T. Michtchenko, S. Ferraz-Mello, Dynamics of two planets in 2:1 mean-motion resonance, Cel. Mech. and Dynamical Astronomy 89 (2004) 201-234.
\bibitem{callegarijr2} N. Callegari Jr, S. Ferraz-Mello, T. A. Michtchenko, Dynamics of two planets in the 3/2 mean-motion resonance: application to the planetary system of the pulsar PSR B1257+12, Celestial Mechanics and Dynamical Astronomy 94 (2006) 381-397.
\bibitem{ferraz} S. Ferraz-Mello, T. A. Michtchenko, C. Beaugé, N. Callegari Jr., Extrasolar Planetary Systems, in: R. Dvorak et al. (Eds.), Chaos and Stability in Exrasolar Planetary Systems, Lectures Notes in Physics, 683, 2005, pp. 219-271.




\end{thebibliography}
\end{document}